# DISCOVERING ULTRA–HIGH-ENERGY NEUTRINOS THROUGH HORIZONTAL AND UPWARD $\tau$ AIR SHOWERS: EVIDENCE IN TERRESTRIAL GAMMA FLASHES?

D. Fargion[1,2]
Dipartimento di Fisica, Università degli Studi di Roma "La Sapienza," Piazza Aldo Moro 2, 00185 Rome, Italy; daniele.fargion@roma1.infn.it

## ABSTRACT

Ultra–high-energy (UHE) neutrinos $\nu_\tau$, $\bar{\nu}_\tau$, and $\bar{\nu}_e$ at PeV and higher energies may induce $\tau$ air showers whose detectability is amplified millions to billions of times by their secondaries. We considered UHE $\nu_\tau$-$N$ and UHE $\bar{\nu}_e$-$e$ interactions underneath mountains as a source of such horizontal amplified $\tau$ air showers. We also consider vertical upward UHE $\nu_\tau$-$N$ interactions (UPTAUs) on Earth's crust, leading to UHE $\tau$ air showers or interactions at the horizon edges (HORTAUs), and their beaming toward high mountain gamma, X-ray, and Cerenkov detectors, and we show their detectability. We notice that such rare upward $\tau$ air showers, UPTAUs and HORTAUs, may even hit nearby balloons or satellites and flash them with short diluted gamma bursts at the edge of the *Compton Gamma Ray Observatory* detection threshold. We suggest the possibility of identifying these events with recently discovered (BATSE) terrestrial gamma flashes (TGFs), and we argue for their probable UHE $\tau$–UHE $\nu_\tau$ origin. From these data, approximated UHE $\nu_\tau$ fluxes and $\Delta m_{\nu_\mu\nu_\tau}$ lower bounds are derived. Known X-ray, gamma, and TeV active Galactic and extragalactic sources have been identified in most TGF arrival directions. Maximal EGRET activity in the Galactic center overlaps with the maximal TGF flux. The UHE cosmic-ray (UHECR) Akino Giant Air Shower Array anisotropy at $10^{18}$ eV also shows possible correlations with TGF events. The unique UHECR triplet in AGASA clustering, pointing toward BL Lac 1ES 0806+524, finds within its error box a corresponding TGF event, BATSE trigger 2444. Finally, a partial TGF Galactic signature, combined with the above correlations, suggests an astrophysical $\tau$ origin of TGF events.

*Subject headings:* cosmic rays — elementary particles — instrumentation: detectors

*On-line material*: color figures

## 1. INTRODUCTION: UHE NEUTRINOS

Ultra–high-energy (UHE) neutrinos of astrophysical origin are waiting to be observed; above tens of TeV, UHE $\nu$'s might overcome the nearby signals of dominant, noisy, secondary atmospheric neutrinos. Being their parent's primary smeared charged particles, such atmospheric neutrinos do not lead to any UHE astrophysics. Present and future underground cubic kilometer detectors are looking for the muon-penetrating tracks to spatially associate remarkably persistent astrophysical sources (active galactic nuclei [AGNs], supernovae [SNe], microquasars) or the rarest gamma-ray burst (GRB) event. Downward muons, the secondaries of air showers, are dominating and polluting the downward vertical signals; upward muons from UHE neutrinos $\nu_\mu$ and $\bar{\nu}_\mu$ at low energies (below TeV) are again polluted by atmospheric neutrinos; higher energy neutrinos $\nu_\mu$ and $\bar{\nu}_\mu$ above $10^{13}$ eV may better probe the astrophysical neutrino, but such upward neutrinos unfortunately are more and more suppressed by Earth's opacity. Upward $\tau$ neutrinos, to be discussed here, are less opaque, but at $10^{13}$–$10^{14}$ eV they leave shorter tracks and are less detectable in cubic kilometer detectors. Therefore, the best strategy in underground detectors that we imagine considers horizontal underground arrays. For this reason, we strongly suggest the construction of wider cubic kilometer ($\sim 4$ km radius) and narrower (0.2 km depth) disklike arrays underground, finalized mainly to horizontal UHE astrophysical neutrinos of both of the heaviest leptons.

Indeed, upward and horizontal $\tau$ air showers emerging from Earth's crust or mountain chains, as discussed in this paper, are, in our opinion, the most powerful signals of UHE neutrinos $\nu_\tau$, $\bar{\nu}_\tau$, and $\bar{\nu}_e$ at PeV and higher energies. The multiplicity in $\tau$ air shower secondary particles, $N_{opt} \simeq 10^{12}$ $(E_\tau/\text{PeV})$, $N_\gamma(\langle E_\gamma \rangle \sim 10 \text{ MeV}) \simeq 10^8 (E_\tau/\text{PeV})$, $N_{e^-e^+} \simeq 2 \times 10^7 (E_\tau/\text{PeV})$, and $N_\mu \simeq 3 \times 10^5 (E_\tau/\text{PeV})^{0.85}$, makes their discovery easy. UHE $\nu_\tau$'s and $\bar{\nu}_\tau$'s, following Super-Kamiokande discovery (as well as the most recent Sudbury Neutrino Observatory [SNO] solar neutrino evidence for all flavor mixing; $\nu_\mu \leftrightarrow \nu_\tau$), should be as abundant as $\nu_\mu$'s and $\bar{\nu}_\mu$'s. Also, antineutrino electrons, $\bar{\nu}_e$'s, near the Glashow $W$ resonance peak, $E_{\bar{\nu}_e} = M_W^2/2m_e \simeq 6.3 \times 10^{15}$ eV, may generate $\tau$ air showers even in the absence of any flavor mixing.

The upward $\tau$ air showers (UPTAUs) and the horizontal $\tau$ air showers (HORTAUs) are analogous to the "double bangs" (Learned & Pakvasa 1995) in underground neutrino detectors. The novelty of the present "one bang in (the rock, Earth's crust)–one bang out (the air)" lies in the huge density of the rock compared to the atmosphere, the self-triggered explosive nature of $\tau$ decay in flight, and its consequently huge amplified air shower signal (at a characteristic distance of a few kilometers) with respect to the unique muon track in cubic kilometer detectors.

Following the remarks of the anonymous referee, one should be reminded that (1) looking for air showers behind mountains has a long history; there are unpublished experimental proposals by A. Abashian in the late 1980s (possibly also seeking new physics as in the Stanford Linear Accelera-tor Center beam dump; see, e.g., Bjorken et al. 1984), (2) the idea of looking for upward air showers from mountain tops was also suggested by T. Bowen from the University of Arizona and discussed at several International Cosmic-Ray Conferences and deep underwater muon and neutrino detector (DUMAND) meetings, and (3) the idea of looking from satellites was pioneered by the Orbiting Wide-Angle Light collector (OWL) collaboration. Because of the lack of refereed literature, it is difficult (actually for the author, it has been impossible) to reference any of this material. However, the OWL subjects are quite different because they concerned only terrestrial atmospheric UHE cosmic-ray (UHECR) and terrestrial atmospheric neutrino showering and not upward UHE $\tau$ air showers born in solid rock (mountains, Earth's crust), whose UPTAUs and HORTAUs are discussed for the first time (Fargion 1997; Fargion, Aiello, & Conversano 1999a) and in detail below. A long list of related articles about the role of $\tau$ were published after our earliest submission and are summarized in footnote 5.

---

[1] Instituto Nazionale di Fisica Nucleare, Piazza Caprettari, 70, Rome I-00186, Italy.
[2] Technion Institute, Engineering Faculty, Haifa, Israel.

The vertical upward $\tau$ air showers (small arrival nadir angle) may occur preferentially at low energies nearly transparent to Earth ($E_\nu \sim 10^{15}$–$10^{16}$ eV). The oblique $\tau$ air showers (whose arrival directions have a large nadir angle) may also be related to higher energy $\nu_\tau$ or $\bar{\nu}_\tau$ nuclear interactions ($E_{\bar{\nu}_\tau} \geq 10^{17}$–$10^{19}$ eV). Indeed, these horizontal upward UHE $\nu_\tau$'s cross a smaller fraction of Earth's volume, and consequently, they suffer less absorption toward the horizon. Moreover, the consequently ultrarelativistic ($E_{\bar{\nu}_\tau} \geq 10^{17}$–$10^{19}$ eV) $\tau$'s may travel in the atmosphere for a few or even hundreds of kilometers with no absorption before the decay to the detector located at a distance of a few (or hundreds) of kilometers. On the contrary, the horizontal gamma, electron pair, and muon showers from primary (downward, nearly horizontal) UHECR protons are severely suppressed ($\leq 10^{-3}$) after crossing ($\geq 2 \times 10^3$ g cm$^{-2}$) the slant depth or, equivalent, at 1 atm ($\geq 16$ km) of the sea-level horizontal atmosphere target (Ave et al. 2000; Cillis & Sciutto 2001). This opacity will lead to a wide ($\sim 10°$) angle between the downward horizontal arrival UHECR above the horizon and the HORTAUs below the same horizon edges. From balloons this angle size will be comparable, while from the high satellite quota it will be as narrow as 1°. This implies a need to discriminate HORTAUs from horizontal high-altitude showers (HIAS; Fargion 2001a). Upward UHE $\nu_\tau$-$N$ interactions on Earth's crust at the horizontal edge and from below and their consequent upward UHE $\tau$ air showers beaming toward high mountains, airplanes, balloons, and satellites should flash gamma, muon, X-ray, and Cerenkov lights toward detectors. Such upward $\tau$ air showers may already hit a nearby satellite, such as the *Compton Gamma Ray Observatory* (*CGRO*), flashing them with short, hard, diluted gamma bursts at the edge of the BATSE (and future *GLAST*) threshold. The $\tau$ air shower may test the UHE neutrino interactions, leading to an additional finetuned test of new TeV gravity physics (by extra dimensions) in both mountain valleys and in upward showers (see Fig. 1 and Appendix C). Any mountain chain acts as a clever multifilter: (1) as a screen of undesirable common horizontal UHECRs (electromagnetic shower, secondary Cerenkov photons, and muons), (2) as a dense calorimeter for UHE $\nu_\tau$'s, $\bar{\nu}_\tau$'s, and $\bar{\nu}_e$'s, (3) as a distance meter target correlating the $\tau$ birthplace and its air shower opening with its most probable energy, (4) as a unique source, through $\tau$ hadronic showering, of the horizontal and deep muon bundle source, and (5) as a calorimeter not opaque to energetic ($10^{16}$–$10^{20}$ eV) UHE neutrinos. Because of the different UHE neutrino interactions with energy and flavors, it will be possible to estimate, by stereoscopic, directional, and time-space structure signatures, the spatial air shower origination and its probable original energy. We also suggest here the possibility of the discovery of UHE $\tau$'s by observing the UPTAUs and HORTAUs, arriving from tens, hundreds, and thousands of kilometers away (near the horizontal edges), from high mountains, high balloons, and satellites; such UHE $\tau$'s created within a wide (tens of thousands to millions of square kilometers wide and hundred of meters deep in Earth's crust) target would discover UHE $\nu_\tau$ and $\bar{\nu}_\tau$ neutrinos at PeV to EeV energies and above, just within the mysterious Greisen-Zatsepin-Kuzmin (GZK) frontiers. The discovery will need capable gamma, optical, and muon bundle detectors sampling a wide-angle view well within the present technology. We show that the UPTAU-HORTAU signal is at the edge of detection of the *CGRO*, namely, BATSE. These signals might already be recorded in terrestrial gamma flashes (TGFs): their gamma beamed fluence of UPTAUs and HORTAUs is just comparable to TGF fluences. Indeed, their observed angular and directional distributions within the Galactic plane (considered in the present article), their clustering toward known active sources (the Galactic center, Crab, M17), and the unique UHECR triplet in the Akino Giant Air Shower Array (AGASA) clustering (toward BL Lac 1ES 0806+524, finding within its error box a corresponding TGF event, BATSE trigger 2444) are very suggestive of their UHE $\nu_\tau$ astrophysical origin. We also note that two of the four additional UHECR AGASA doublets (pointing to BL Lac 2EG J0432+2910 and TEX 1428+370) find two well-correlated TGF events (BATSE triggers 5317 and 2955).

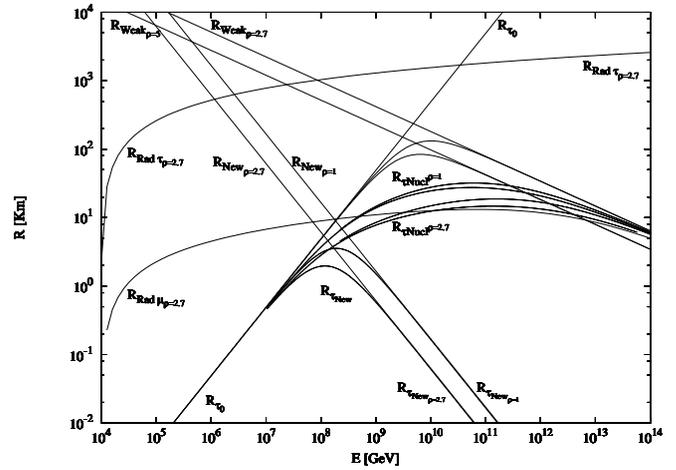

Fig. 1.—Ranges of $\tau$ as a function of the energy for the $\tau$ lifetime $R_{\tau_0}$ (eq. [C1]), the $\tau$ radiation range $R_{\mathrm{Rad}_\tau}$ (eq. [C3]), the $\mu$ radiation range $R_{\mathrm{Rad}_\mu}$, the $\tau$ electro-weak interaction range $R_{\mathrm{Weak}_\rho}$ (eq. [C6]), two densities of $\rho_r$ (2.7 for superficial terrestrial crust and 5 for inner Earth volume), and the range in combined form by their inverse length sum below the radiation range $R_\mu$ for muons. In the central corner a twin line bump for two approximations in eqs. (C4) and (C5) shows the combined more restrictive photonuclear radiation lengths $R_{\tau \mathrm{Nucl}_\rho}$ for either water (*thick curve*) or rock; a possible new TeV physics (extradimensional gravity) and new $\nu$ and $\tau$ range interactions are shown in the corresponding lengths as in eq. (C7), $R_{\tau \mathrm{New}_\rho}$, combined with $R_{\tau_0}$. See Appendix C.

## 1.1. *Neutrino Mass and Mixing*

We wish to remind our readers of the key role of neutrino mass in high-energy astrophysics and cosmology: following the fundamental Gell-Mann (1962) and Cabibbo (1963) discovery of flavor mixing for quarks, Pontecorvo (1963) promptly suggested a neutral lepton mass mixing responsible for light neutrino flavor changing. The obvious need for a nonvanishing neutrino mass led to a wide chain of consequences in astrophysical and cosmological problems, ranging from the role of dark matter (hot dark matter) in the closure of the universe (Pontecorvo 1961) to galaxy formation and dark halos (Pontecorvo 1963; Zeldovich et al. 1980). Therefore, the (probable) heaviest neutral lepton neutrino $\tau$, $\nu_\tau$, may deeply influence the same cosmological history and future evolution with its mass. In particular, its early gravitational clustering in Galactic halos may offer a very efficient gravitational seed for baryonic clustering (Fargion 1983). Other fundamental astrophysical and cosmology links are well known (Weinberg 1972; Dolgov & Zeldovich 1981).[3]

In particular, the solar neutrino puzzle that is finally finding a definitive solution from GALLEX and SNO data calls for mixing and lightest neutrino square mass splitting of $\sim 10^{-3}$ eV$^2$; the supernova neutrino fluxes, observed in SN 1987A, confirmed stellar explosion models that constrain its mass below tens of eV; the Super Kamiokande–MACRO atmospheric neutrino asymmetry implies a light but nonnegligible $\geq 4 \times 10^{-2}$ eV neutrino mass; the hot dark matter in the Galaxy or in galactic cluster halos, the same cosmological dark mass, calls for a $\sim$eV neutrino mass; in this frame, light $\nu$'s, while forming diluted dark Galactic or cluster halos, may offer an ideal calorimeter to UHE $\nu$'s at GZK energies, solving the UHECR puzzle at ZeV energies (Fargion & Salis 1997; Fargion, Salis, & Mele 1999b; Yoshida, Sigl, & Lee 1998; Weiler 1999) and possibly also the TeV-IR cutoff puzzle (Fargion et al. 2001a, 2001b). These interactions may discriminate the neutrino mass (Fargion et al. 2001a, 2001b). Because of the observed mass hierarchy in the lepton sector, neutrino $\tau$'s, associated with the heaviest known charged lepton (Perl et al. 1975), may well be the heaviest and, therefore, dominant ones. Moreover, neutrino mass may be of either Majorana or Dirac nature, leading to different elementary particle behaviors and early nucleosynthesis evolution (Pontecorvo 1963; Fargion & Shepkin 1984). The same neutrino mass implies the presence (by Lorentz boost) of the "sterile" right-handed partner, whose role may be important in the early universe (Fargion 1981b; Antonelli, Konoplich, & Fargion 1981) and in the highest energy astrophysics. The different $\nu_L$-$\nu_R$ interactions and the different thermal evolutions may lead to different thermal neutrino populations. Just to emphasize the roles of $\nu$ mass with respect to massless gravitons, we remind the reader of the important case of an SN MeV neutrino burst arrival being slowed by its mass relativistic flight and its delayed arrival from far SN (Galactic or, even better, extragalactic) events with respect to the massless gravitational waves. The expected time delay between the massless graviton wave burst (by supernova quadrupole emission at distance $L$) and the $\nu_e$ neutronization neutrino burst, whose timescale is close to 0.4 ms, will be easily detectable, leading to an additional test of the elusive mass detection: $\Delta t \sim 50$ s $(E_\nu/5 \text{ MeV})^{-2}(m_\nu/5 \text{ eV})^2(L/\text{Mpc})$. In particular, this delay, after flavor oscillations, must in principle be already detectable for the known minimal mass spread (0.05 eV) between muon and $\tau$ flavor masses for any SN located as far as Andromeda (Fargion 1981a). Present neutrino detectors such as SK-SNO-Amanda may correlate with gravitational detectors such as the VIRGO and LIGO detectors measuring bare neutrino masses. The different neutrino mass eigenvalues will also lead to detectable tiny "double" (or even "triple") neutronization neutrino bangs preceded by the sharp gravitational wave burst. The later thermal neutrino burst, while being 10 times more energetic than the neutronization burst, has a longer duration, which cannot be used as well as a good time trigger. The neutrino mass and flavor mixing that have been discovered opened the road to UHE GZK neutrinos and to UHE $\nu_\tau$ astrophysics that can possibly be detected by UPTAUs and HORTAUs.

## 1.2. *The UHE $\bar{\nu}_e$, $\nu_\tau$, $\bar{\nu}_\tau$, and $\tau$ Lengths*

The UHE $\bar{\nu}_e$'s, $\bar{\nu}_\mu$'s, and $\nu_\mu$'s are expected UHECR ($\gtrsim 10^{16}$ eV) secondary default products near AGNs, galactic microquasars, and supernova remnants (SNRs) by common photopion decay relics by optical photons nearby the source (pulsars [PSRs], AGNs; $p + \gamma \to n + \pi^+$, $\pi^+ \to \mu^+ \nu_\mu$, $\mu^+ \to e^+ \nu_e \bar{\nu}_\mu$) and by common UHE proton-nuclear interactions. In this case their source directionality is frozen and conserved. Also, UHE $\nu$ secondaries may be very rare photopion products during the EeV cosmic-ray propagation and interaction in the diffused Galactic lights and inside the Galactic plane gases. In this (rare) case they may lose their primordial source directionality. However, the small but remarkable EeV anisotropy and Galactic directionality observed by AGASA (very possibly related to UHE EeV neutron components) might also trace its presence in UHE neutrinos. We discuss and show the probable UHECR-TGF correlation. We remind the reader that the $\bar{\nu}_e$ signal at the Glashow resonance peak does not depend on any $\nu_\tau$ production or any $\nu_\mu$-$\nu_\tau$ flavor mixing, but it is also source of HORTAUs behind a mountain chain. Neutrino $\tau$'s may be easily produced because of the large Galactic (kiloparsecs) and extreme cosmic (megaparsecs) distances with respect to the neutrino oscillation distance:

$$L_{\nu_\mu\text{-}\nu_\tau} = 4 \times 10^{-3} \text{ pc} \left(\frac{E_\nu}{10^{16} \text{ eV}}\right) \left[\frac{\Delta m_{ij}^2}{(10^{-2} \text{ eV})^2}\right]^{-1}. \quad (1)$$

Such UHE $\nu_\tau$'s and $\bar{\nu}_\tau$'s as well as the UHE $\bar{\nu}_e$'s near a narrow energy resonant $W$ peak, $E_{\bar{\nu}_e} = M_W^2/2m_e = 6.3 \times 10^{15}$ eV, may interact on Earth (the calorimeter), leading to UHE $\tau$'s, which are mostly absorbed by the same planet. However, rare upward UHE $\tau$'s, born by $\nu_\tau$ and $\bar{\nu}_\tau$ nuclear interactions (or rare $\bar{\nu}_e$-$e$ interactions near the upward Earth's surface) may escape outside into the air, where they may spontaneously decay, triggering upward vertical, oblique, or near-horizontal $\tau$ air showers. The vertical ones (with a small nadir angle), UPTAUs, occur preferentially at low energies nearly transparent to Earth ($E_\nu \sim 10^{15}$–$10^{16}$ eV). The oblique $\tau$ air showers, whose arrival directions have large nadir angles, are related mainly to higher energy $\nu_\tau$ or $\bar{\nu}_\tau$ nuclear interactions ($E_{\bar{\nu}_\tau} \geq 10^{17}$–$10^{20}$ eV). Indeed, these horizontal upward UHE $\nu_\tau$ HORTAUs cross a smaller fraction of Earth's volume, and consequently, they suffer less absorption toward the horizon. These HORTAUs, complementary to the PeV $\bar{\nu}_e$ and PeV to tens of PeV $\nu_\tau$ and $\bar{\nu}_\tau$ UPTAUs, are leading to horizontal air showers underneath high mountains and deep valleys and should also be discovered using high ($\gtrsim$ kilometers) mountain observatories hunting for secondary shower Cerenkov lights, muon bundles, and directional electron gamma shower bursts toward Earth. The maximal observational distances may reach $\sim$110 km $(h/\text{km})^{1/2}$ toward the horizon, corresponding to a remarkably finetuned UHE $\tau$ energy, $\sim 2 \times 10^{18}$ eV $(h/\text{km})^{1/2}$. Therefore, we propose the consideration of the nearly horizontal detection of such upward showers from high mountains to test this highest $\nu_\tau$-$\bar{\nu}_\tau$ energy window, which, it should be noted, is almost "blind" to Glashow UHE $\bar{\nu}_e$ fluxes. The comparison of upward showers with the horizontal $\tau$ showering underneath mountains, also made up (20%) of $\bar{\nu}_e$'s at $6.3 \times 10^{15}$ eV, would constrain or even measure the arrival neutrino flavor mixing parameters.

---

[3] The possible presence of an additional fourth neutrino (and quark) family, while unnecessary, is allowed near $M_Z/2$, and it may lead to very exciting astroparticle consequences and observational possibilities in underground detectors (DAMA) and the Galactic GeV EGRET diffused halo (Fargion et al. 1995, 1996, 1998, 2000; Golubkov et al. 1999).

The same $\tau$ upward air showers, UPTAUs and HORTAUs, may penetrate high altitudes, leaving rare beamed upward gamma shower bursts whose sharp ($\sim$hundreds of microseconds) time structure and hard bremsstrahlung ($\gtrsim 10^5$ eV) spectra may hit high-altitude planes near balloons or terrestrial satellites. Here we claim that such gamma upward events originated in $\tau$ air showers, producing gamma bursts at the edge of the *CGRO*-BATSE sensitivity threshold. In particular, we argue that very probably such tiny upward gamma events have already been detected since 1991 April as unexpected sharp ($\lesssim 10^{-3}$ s) and hard ($\gtrsim 10^5$ eV) BATSE gamma triggers originating from Earth and identified, consequently, as TGFs. However, since then (Fishman et al. 1994) TGF understanding of the presently known 75 records (there were over nearly 8000 BATSE triggers in last decade of *CGRO* life and three additional ones during the 2000 yr activity) is based on an unexpected and mysterious high-latitude lightening of a geophysical nature (the so-called sprites or blue jets). We do not agree with this interpretation. While blue jets might, in principle, be triggered by HORTAU air showers in the atmosphere (a giant "Wilson" room amplifier), we believe that they are not themselves the real cause of observed TGFs. We notice that among the 75 records only the details of 47 are published, while 28 TGF events are still unpublished. Their release may be a decisive step in understanding the suggested TGF-UPTAU connection.

## 2. USING $\tau$ AIR SHOWERS TO DISCOVER UHE $\nu$'s: UHE $\tau$ DECAY CHANNELS

The $\tau$ air shower morphology would reflect the rich and structured behaviors of $\tau$ decay modes. Indeed, let us label the main "eight finger" UHE decay channels (hadronic or electromagnetic) and the consequent air shower imprint with the corresponding probability ratio as shown in Table 1.

This complex air shower mode would exhibit different interaction lengths in the air at 1 atm ($\sim$300 m for electromagnetic interaction length, 500 m for the hadronic interaction length, or more precisely, 800 m for $\tau$ pions secondaries). The consequent air shower statistics will also reflect these imprint multichannel modes in their energy and structured time arrival to detectors underneath mountains on planes, balloons, or satellites. These channels may also possibly be reflected in observed TGFs.

### 2.1. *HORTAUs underneath Mountain Chains*

HORTAUs may be a key signal of UHE neutrino astrophysics. However, UHE muon and $\tau$ tracks above few tens of PeVs, in a cubic kilometer icewater detector, leave kilometer length traces that are difficult to distinguish: a shower in water might be either a catastrophic bremsstrahlung muon interaction or a $\tau$ decay in flight. The tens of kilometers range of UHE $\tau$'s at horizons in limited underground (cubic kilometer) detectors is not distinguished from PeV $\mu_S$'s. In order to recognize with no ambiguity each lepton nature, we propose here a new detector based on UHE $\tau$ air showers able to filter and reveal $\tau$ leptons (Fargion et al. 1999a). We suggest for this aim that a deep valley such as the Grand Canyon, the famous Death Valley, or the nearby Inyo-White and Whitney mountains in Nevada, as well as glacial valleys in the Alps or the fjords in Scandinavian regions be considered. It will be possible to observe UHE neutrinos by their contained horizontal showers in huge air volumes. The deep rock walls play the same role as the target beam dump for UHE neutrinos as well as filtering atmospheric cosmic rays. Moreover, the same kilometer volume sizes act as a filter, avoiding random atmospheric muon decay but allowing the UHE $\tau$ decay at the PeV band. The same rock may increase the target matter (with respect to the horizontal shower in air as in the Auger experiment) by nearly 2 orders of magnitude. At a large depth ($\geq$1 km) and at horizontal angles (60°–90°) at energies above 1000 TeV the atmospheric muon secondaries (crossing 2 km or more of rock matter) will be negligible, while a primary neutrino may interact in the rock, leading to secondary leptons: only muon and $\tau$ tracks are long enough to emerge often from the mountain wall rock. At the $W$ pole ($E = 6.3$ PeV), the UHE secondary muon will decay, in empty space, at a distance of $10^7$ km, while the corresponding $\tau$ is at only 70 m. Such muons may poorly interact by bremsstrahlung in air beyond $10^4$ km; lower energetic muons that originated in the mountains at GeV band windows will radiate at a negligible level. Therefore, PeV muons in air will produce sterile single tracks, while the corresponding $\tau$'s should decay, leading to a copious air shower originating within 50 m of the mountain walls. These bounded horizontal atmospheric showers originating within the deep valley may be observed by either Cerenkov detectors inside the cubic kilometer volumes or, more economically, just monitoring above the valley the atmospheric scintillation traces and their fluorescent signals in the dark nights. Our first estimates for $\tau$ horizontal showers are based on an interaction volume defined for an ideal example as follows: the Argentier Alps (Fig. 2) chain distance ($D \simeq 10$ km), the characteristic height ($\sim 1$ km), the UHE $\nu_\tau$-$\tau$ energies ($\sim 3$ PeV), the $\tau$ distance before the decay ($\sim 150$ m), and the total interactive volume ($V \simeq 4.5$ km$^3$; water equivalent). The air shower tail may also be observed as a gamma burst. The $\tau$ air shower volume is therefore observable within a narrow beamed cone (Moliere radius $\sim 80$ m; distance $\sim 5$ km; $\Delta\theta \sim 1°$; $\Delta\Omega \sim 2 \times 10^{-5}$), and it is reduced to an effective volume $V_{\mathrm{eff}} \simeq 9 \times 10^{-5}$ km$^3$ for each observational detector (Fig. 2). Each one is comparable to roughly twice a Super-Kamiokande detector. We expect, following the AGN SS 91 model of Gandhi et al. (1998), a total rate of $6(\bar{\nu}_e e) + 29(\nu_\tau N) = 35$ UHE $\nu_\tau$ events yr$^{-1}$ km$^{-3}$; at energies above 3 PeV we may expect a total rate of $N_{\mathrm{ev}} \sim 158$ events yr$^{-1}$ in this Argentiere Alps mountain valley (Fig. 2) and at least $3.2 \times 10^{-3}$ events yr$^{-1}$ for each detector. Wider angle acceptance (which is realistic) may lead to an order-of-magnitude increase in the detection rate at $E_\tau \sim 3$ PeV,

$$N_{\mathrm{ev}} \sim 158 \left(\frac{h}{1 \text{ km}}\right)\left(\frac{D}{10 \text{ km}}\right) \text{ events yr}^{-1}. \quad (2)$$

TABLE 1
$\tau$ Air Shower Channels

| Decay | Secondaries | Probability (%) | Air Shower |
|---|---|---|---|
| $\tau \to \mu^- \bar{\nu}_\mu \nu_\tau$ | $\mu^-$ | $\sim$17.4 | Unobservable |
| $\tau \to e^- \bar{\nu}_e \nu_\tau$ | $e^-$ | $\sim$17.8 | One electromagnetic |
| $\tau \to \pi^- \nu_\tau$ | $\pi^-$ | $\sim$11.8 | One hadronic |
| $\tau \to \pi^- \pi^0 \nu_\tau$ | $\pi^-, \pi^0 \to 2\gamma$ | $\sim$25.8 | One hadronic, two electromagnetic |
| $\tau \to \pi^- 2\pi^0 \nu_\tau$ | $\pi^-, 2\pi^0 \to 4\gamma$ | $\sim$10.79 | One hadronic, four electromagnetic |
| $\tau \to \pi^- 3\pi^0 \nu_\tau$ | $\pi^-, 3\pi^0 \to 6\gamma$ | $\sim$1.23 | One hadronic, six electromagnetic |
| $\tau \to \pi^- \pi^- \pi^+ \nu_\tau$ | $2\pi^-, \pi^+$ | $\sim$10 | Three hadronic |
| $\tau \to \pi^- \pi^+ \pi^- \pi^0$ | $2\pi^-, \pi^+, \pi^0 \to 2\gamma$ | $\sim$5.18 | Three hadronic, two electromagnetic |

Because of simple power laws, it is possible to extend to higher energies this estimate of general neutrino fluxes, assuming a neutrino power law $E^{-2}$ whose fluence scales linearly with present AGASA bounds ($3 \times 10^3$ eV cm$^{-2}$ s$^{-1}$ sr$^{-1}$) used for present estimate:

$$N_{\mathrm{ev}} \sim 158 \left(\frac{h}{1 \text{ km}}\right)\left(\frac{D}{10 \text{ km}}\right)\left(\frac{E_\nu}{3 \text{ PeV}}\right)^{0.36} \text{ events yr}^{-1}. \quad (3)$$

Contrary to the UPTAUs to be described in next section, for HORTAUs behind a mountain, there is no neutrino opacity at the PeV–ZeV energy band (see Figs. 1 and 3). A few hundred detectors at distances of hundreds of meters should be located along the valley, and their signal-to-noise ratio would be strong enough. The most abundant $\tau$ air shower signature, at the 5 km air distance atmosphere, corresponding to an $X_0 = 625$ g cm$^{-2}$ radiation length, would be a copious electromagnetic (and often additional hadronic and muonic) air shower whose hard (MeV) spectra have a flux rate of at least $\phi_\gamma \gtrsim 3 \times 10^{-2}$ cm$^{-2}$ s$^{-1}$. A few square meter scintillator would observe tens (or hundreds) of MeV gamma events in a very narrow gamma burst arrival angle and time in a screened detector (from upper electromagnetic showers above the edge of the target mountain; Fig. 2). The burst timing signal will be a few microseconds long, and its directionality would be easily correlated toward the horizontal mountains (Fig. 2). Technical details will be discussed elsewhere.

### 2.2. Upward $\tau$ Air Showers: UPTAUs

As we have shown, the small flux and cross section of UHE $\nu_\tau$'s call for wider and wider target volume. As the OWL experimental proposal has noted, the whole Earth may be the ideal beam dump not only for UHECRs born in air but also for the UHE $\nu_\tau$'s coming up from Earth and their $\tau$ secondaries decaying in space. The UHE $\nu$-$N$ interactions may generate their parental UPTAUs, which will amplify the signal. The Earth opacity at 3 PeV has been evaluated (Gandhi et al. 1998) and corresponds in the Stecker-Salomon model to 5.2 events km$^{-3}$ yr$^{-1}$ for upward muons. The corresponding event birth rates for UHE $\tau$'s (Dutta et al. 2001) are within a $\nu$ flux unity model at $3 \times 10^3$ eV cm$^{-2}$ s$^{-1}$ and are nearly twice as large: $N_{\rm ev} \sim 10$ events km$^{-3}$ yr$^{-1}$. In a first approximation it is possible to show that the Earth volume observable from the top of a mountain at height $h$, because of UHE $\tau$'s at 3 PeV crossing from below, is approximately $V \approx 5 \times 10^4$ km$^3 (h/{\rm km})$ $(E_\tau/3 \text{ PeV})$. The UPTAUs would hit the top of the mountain (Fig. 4). For the same $\tau$ air shower beaming ($\Delta\theta \sim 1°$, $\Delta\Omega \sim 2 \times 10^{-5}$), we now derive an effective volume of $\sim 1$ km$^3$. Therefore, a detector open at a $2\pi$ angle on top of a 2 km high mountain may observe nearly an event every 2 months from below Earth. The gamma signal above a few MeV would be (depending on the arrival nadir angle) between $3 \times 10^{-2}$ (for a small nadir angle) and $10^{-5}$ cm$^{-2}$ s$^{-1}$ at far distances at 3 PeV. Below MeV to tens of KeV, the signal (secondary bremsstrahlung hard X-ray photons) may be 100 times larger. However, because of the variable Earth opacity with nadir angle, the highest UHE $\nu_\tau$ and $\tau$ events would also arrive easily at the horizon, leading to a compensation or even an amplification in the average gamma flux at the horizon edge. At a large nadir angle ($\gtrsim 60°$) where an average Earth density may be assumed ($\langle\rho\rangle \sim 5$), the transmission probability and creation of upward UHE $\tau$'s is approximately

$$P(\theta, E_\nu) = e^{[-2R_{\rm Earth} \cos\theta / R_{\nu_\tau}(E_\nu)]} \left(1 - e^{-[R_\tau(E_\tau)/R_{\nu_\tau}(E_\nu)]}\right). \quad (4)$$

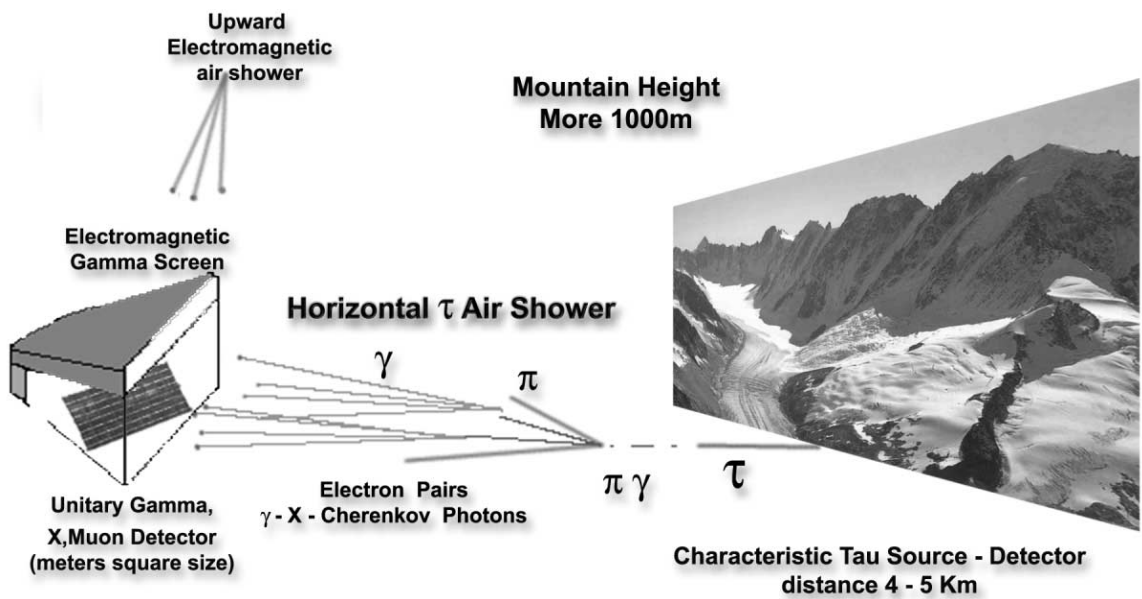

Fig. 2.—Schematic drawing of hypothetical horizontal air showers originating in a mountain chain such as the Alps Argentier mountain chain and the detailed and enlarged scheme of the $\gamma/X$ burst $\tau$ secondaries signal; a shield from downward electromagnetic air showers is also described.

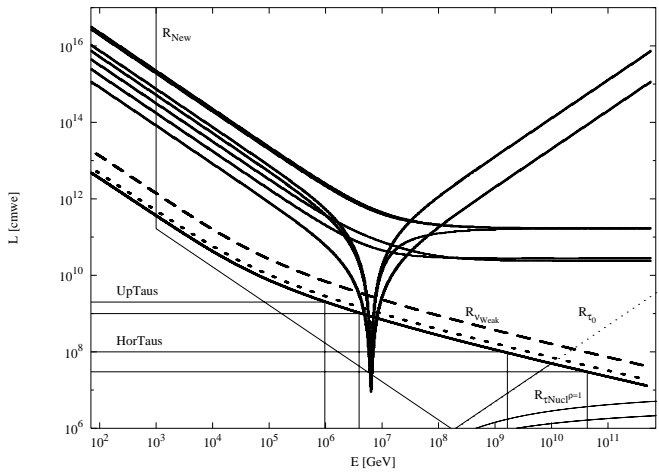

FIG. 3.—UHE $\nu$ ranges $R_\nu$ from $\nu_\tau$-$N$, and $\bar{\nu}_\tau$-$N$ interactions as a function of UHE neutrino energy in Earth; overlapping the $\bar{\nu}_e$-$e$ resonant interactions and ranges leads to PeV HORTAUs. In the right-hand corner below the $R_\tau$ interactions lengths are $R_{\tau 0}$ and the main UHE $\tau$ range $R_{\tau{\rm Nucl}\rho}$, as in Fig. 1, at the same energies in water and (below) in rock. The energy windows for UPTAUs and HORTAUs (from a satellite) are shown. Nevertheless, finetuned UHE $\tau$ energy [$\sim 2 \times 10^{18}$ eV $(h/{\rm km})^{1/2}$] allows all the HORTAU energies. New TeV physics introduces an additional constraining range for both $R_\nu$ and $R_\tau$ as defined by the line $R_{\rm New}$ for $\rho = 1$.

The corresponding angular integral effective volume observable from a high mountain (or balloon) at height $h$ (assuming a final target terrestrial density $\rho = 3$) is

$$V_{\rm eff} \approx 0.3 \text{ km}^3 \frac{\rho}{3} \frac{h}{\rm km} e^{-(E/3 \text{ PeV})} \left(\frac{E}{3 \text{ PeV}}\right)^{1.363}. \quad (5)$$

Because the upward event from UHE $\nu_\tau$'s above 3 PeV leads to $\sim 10$ events km$^{-3}$ yr$^{-1}$, we must expect that an average effective event rate for each year on a top of a mountain ($h \sim 2$ km; Fig. 4) for incoming energy flux $\Phi_\nu$ (in units of cm$^{-2}$ s$^{-1}$ eV), as in Figure 5, is

$$N_{\rm eff} \simeq 8 \frac{\rho}{3} \frac{h}{2 \text{ km}} e^{-(E/3 \text{ PeV})} \left(\frac{\Phi_\nu}{3 \times 10^3}\right) \left(\frac{E}{3 \text{ PeV}}\right)^{1.363}. \quad (6)$$

This rate is quite large; it may be scaled with incoming neutrino fluxes and the expected $\tau$ air shower signals (gamma burst at energies $\gtrsim 10$ MeV) and should be $\phi_\gamma \simeq 10^{-4}$ to $10^{-5}$ cm$^{-2}$ s$^{-1}$, while the gamma flux at $\sim 10^5$ eV or lower energies (from electron pair bremsstrahlung) may be 2 orders of magnitude larger.

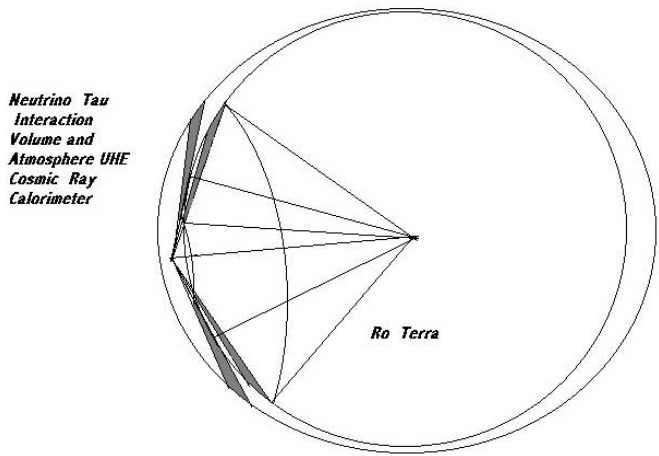

FIG. 4.—Schematic geometrical drawing of the nearly horizontal upcoming air shower cones, from both above and below the Earth's crust and the target of UHE neutrino interaction and, consequently, the nearly horizontal $\tau$ air shower (HORTAU). [*See the electronic edition of the Journal for a color version of this figure.*]

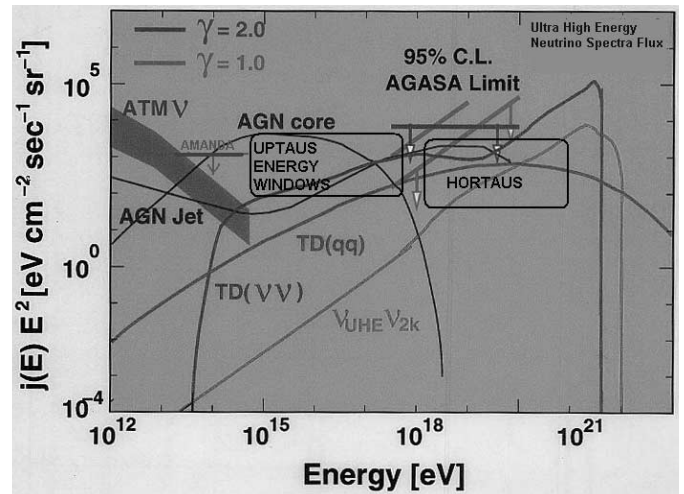

FIG. 5.—Expected UHE neutrinos in UPTAU and HORTAU energy window fluence and recent upper bounds for different UHE neutrino source models following Yoshida (2001). [*See the electronic edition of the Journal for a color version of this figure.*]

A collection of scintillators and Cerenkov light detectors (e.g., CASA-BLANCA), better screened from upward UHECR electromagnetic showers (Fig. 2), may easily discover this UHE (PeV) flux. Detailed schemes will be shown elsewhere. We notice that the horizontal atmosphere (air) target depth (~360 m water equivalent) at the UHE $\tau$ peak of ~$10^{19}$ eV ($R_\tau \sim 500$ km; eq. [C6]) is nearly 1000 times smaller than the corresponding one through the Earth's crust depth (from the horizon), leading to a corresponding suppression factor of 1000 in the $\nu_\tau$-$\tau$ visibility from above (with respect to those arriving from below the horizon; Fig. 4). In summary, the HORTAUs (at GZK) induce air showers 3 orders of magnitude better than horizontal downward $\tau$ air showers.

### 2.3. Horizontal $\tau$ Air Shower from Earth: HORTAUs

The possibility for a HORTAU to reach tangentially high altitudes and hit a satellite is related to (1) the UHE $\nu_\tau$-$N$ interaction probability within the Earth, (2) the UHE $\tau$ track in the rock at the upper crust, (3) the $\tau$ decay in flight distance within the atmosphere, and (4) the radiation length in air and air shower amplification and suppression. The visible Earth surface $S$ from a satellite, like BATSE, at height $h \sim 400$ km is $S = 2\pi R_\oplus [1 - \cos(\theta_c)] \simeq 2\pi h R_\oplus \simeq 4 \times 10^4 \times$ km$^2(h/$km$)$ for $h \ll R_\oplus$, and the consequent effective volume for UHE $\nu_\tau$-$N$ interaction at 3 PeV and HORTAUs beamed within $\Delta\Omega \sim 2 \times 10^{-5}$ rad$^2$, is (note that $\langle\rho\rangle \simeq 1.6$ because 70% of Earth is covered by water)

$$V_{\text{eff}} = V_{\text{TOT}} \Delta\Omega \simeq 60 \text{ km}^3 \left(\frac{h}{400 \text{ km}}\right). \quad (7)$$

The effective volume and the event rate should be reduced at a large nadir angle ($\theta > 60°$) by the atmosphere depth and opacity (for a given $E_\tau$ energy). Therefore, the observable volume may be reduced approximately to within 15 km$^3$, and the expected UHE PeV event rate is

$$N_{\text{ev}} \sim 150 \frac{h}{400 \text{ km}} \text{ events yr}^{-1} \ (E_\tau \geq 3 \text{ PeV}). \quad (8)$$

### 3. UPTAUs AND HORTAUs TOWARD SATELLITES: TGF EVENTS IN BATSE DATA

Let us estimate the possible role of UPTAUs (and HORTAUs) in triggering a TGF. The present rate of observed TGFs (low threshold and hard channel trigger setup) is at best much lower (~a factor of 10) than the above formula: it may well be possible that the usual BATSE trigger is suppressing and hiding this rate. Otherwise, tens of PeV UHE $\nu_\tau$'s are the TGF event sources at the BATSE sensibility threshold. Therefore, a small (a factor of 3–5) exponential suppression, as in the above equation, may reduce the $N_{\text{ev}}$ to the observed TGF rate, while at the same time it may slightly increase their intensities. Moreover, HORTAUs at tens of EeV (Fargion 2001a, 2001b, 2001c) may also lead to rare upward events at a rate comparable to TGF events. The last $\tau$ air shower traces of 3 PeV are mainly hard ($10^5$ eV or above) bremsstrahlung gamma photons of the last air shower electron pairs whose approximated number flux is comparable to

$$N_\gamma \simeq \frac{E_\tau}{\langle E_\gamma \rangle} \sim 3 \times 10^{10} \left(\frac{E}{3 \text{ PeV}}\right). \quad (9)$$

The atmosphere opacity may reduce the final value to at least $\frac{1}{3}$: $N_\gamma \sim 10^{10}$. The expected X-ray–gamma flux at large 500 km distances is diluted even within a beamed angle $\Delta\Omega \simeq 2 \times 10^{-5}$, leading to nearly $\Phi_\gamma \sim 10^{-2}$ photons cm$^{-2}$. The characteristic gamma burst duration is roughly defined by $L/c \sim$ few milliseconds, in agreement with the observed TGF events (see Appendix A). The consequent TGF gamma burst flux is $\Phi_\gamma S \sim 10^2$ events, just comparable with TGF observed events (see Appendix B). The HORTAUs, while being more energetic (a factor of ~$10^3$ for the same $\nu$ energy fluence), are rarer by the same factor (for equal $\nu$ energy fluence) and by a smaller arrival angle (2 orders of magnitude) as well as being diluted by the longer tangential distances (a factor of ~25); however, most of these suppressions are well recovered by a much higher and efficient $\nu$ cross section at GZK energies, longer $\tau$ tracks, possible rich primary spectra, and higher HORTAU $\Phi_\gamma$ fluence, making them complementary or even comparable to UPTAU signals. The bremsstrahlung spectra is as hard as the observed TGF spectra. The possible air shower time structure may reflect the eight different $\tau$ decay channels (mainly hadronic and/or electromagnetic ones). The complex interplay between UHE $\nu$ interaction with nuclear matter superimposed on $\bar{\nu}_e$-$e$ interactions is shown in Figure 3. The extremely narrow energy window where the $\bar{\nu}_e$-$e$ rate is comparable to $\nu_\tau$-$N$ while being transparent to Earth makes the UPTAU-HORTAU-TGF connection unrelated to $\bar{\nu}_e$-$e$ resonant $W^-$ events possible only in HORTAUs underneath a mountain. The characteristic interaction regions responsible for UPTAUs and HORTAUs is within a narrow energy band described in Figures 3 and 5. The peculiar $\nu_\tau$-$N$ interaction (Fig. 6) departing from parton model would lead to a less restrictive UHE $\nu_\tau$ Earth opacity and a more abundant vertical UPTAU-TGF event rate at higher energy; the TGF data do not support such a large flux variability, and therefore, it might moderately favor the narrow energy window (PeV to a few tens of PeV) model constrained by partons (Fig. 3) or the EeV energy window for HORTAUs. Indeed, the TGF data, collected from the NASA BATSE archive and described in Table 3, are located in a celestial map with

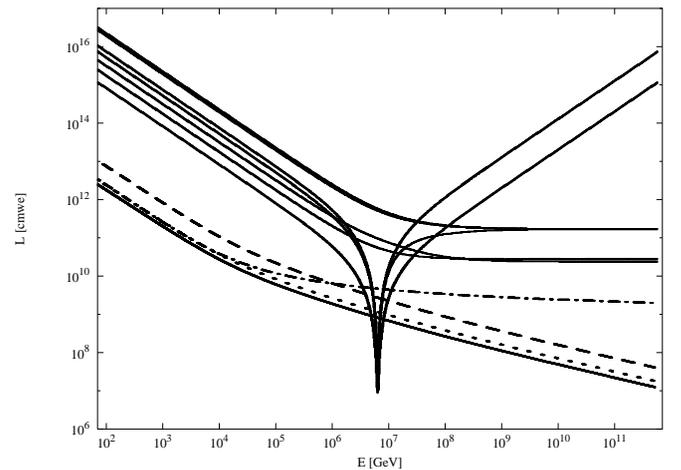

Fig. 6.—As in Fig. 3, following Gandhi et al. (1998) UHE neutrino ranges as a function of UHE neutrino energy in Earth with overlapping $\bar{\nu}_e$-$e$, $\nu_\tau$-$N$, and $\bar{\nu}_\tau$-$N$ interactions; different weaker UHE neutrino interaction models are also present, leading to less terrestrial opacity to both UHE PeV as well as EeV $\nu_\tau$'s.

TABLE 2
X-Ray–Gamma Sources

| Name | R.A. (arcsec) | Decl. (arcsec) | Source Type | TeV Source |
|---|---|---|---|---|
| Crab | 83.52 | 22.19 | Anti–Galactic center PSR | A |
| Geminga | 98.48 | 17.77 | Anti–Galactic center gamma PSR | |
| PSR 1706−44 | 257.4 | −44.52 | Galactic center gamma PSR | A |
| PKS 0528 | 82.73 | 13.53 | Anti–Galactic center AGN | |
| PKS 2155−304 | 329.72 | −30.22 | AGN | B |
| Cyg X-3 | 308.11 | 40.96 | PSR | |
| 3C 273 | 187.28 | 2.05 | Brightest nearby gamma quasars | |
| 3C 279 | 194.05 | −5.79 | Brightest gamma quasars | |
| Mrk 421 | 166.11 | 38.21 | Unmatched TeV AGN sources | A |
| Vela X-1 | 135.28 | −40.55 | Galactic gamma PSR | B |
| Virgo | 187.5 | 13.2 | Nearby Galactic clusters | |
| G 21.5 | 278 | −10 | AGN | |
| Perseo | 49.96 | 41.53 | Nearby Galactic clusters | |
| Coma | 194 | 28 | Nearby Galactic clusters | |
| Flys' Eye | 87 | 48.1 | Unmatched UHECR sources | |
| Mrk 501 | 253.47 | 39.76 | Unmatched TeV AGN sources | A |
| Cen A | 201.37 | −43.02 | Nearby AGN | |
| GRS 1915−105 | 289.33 | 10.55 | Microquasar | |
| SCO X-1 | 244.98 | −15.64 | X-ray PSR | |
| $\rho$ Oph | 247.03 | −24.54 | X-ray PSR | |
| G.A. 1740-7-2942 | 266.01 | −29.72 | Galactic center gamma-TeV PSR | |
| Mrk 279 | 208.26 | 69.31 | AGN | |
| GRO J2250−13 | 342.5 | −13 | gamma PSR | |
| SN 1006 | 225 | −41.5 | Nearby Galactic center SNR and PSR | B− |
| M31 | 10 | 40 | M31 Andromeda | |
| PKS 0235+164 | 39.66 | 16.61 | AGN | |
| 2EG J0239+2818 | 40.09 | 28.21 | Unidentified EGRET source | |
| 2EGS J0500+5902 | 75.15 | 59.04 | Unidentified EGRET source | |
| 3C 454.3 | 343.49 | 16.13 | Quasars or AGN | |
| J0319+2407 | 49.76 | 24.12 | Quasars or AGN | |
| Her X-1 | 254.46 | 35.34 | Unmatched sources | |
| PKS 1622−297 | 246.53 | −29.86 | Galactic center PSR | |
| PKS 1959+650 | 30 | 65.08 | Quasars or AGN | B− |

Note.—The "TeV Source" column indicates the most prominent TeV sources defind by the categorization of Weeks.

their corresponding error boxes (Fig. 7). They are more readable, after a few error bar calibrations, in a Galactic map over the diffused GeV EGRET gamma background signal (Fig. 8). One should note the surprising clustering of

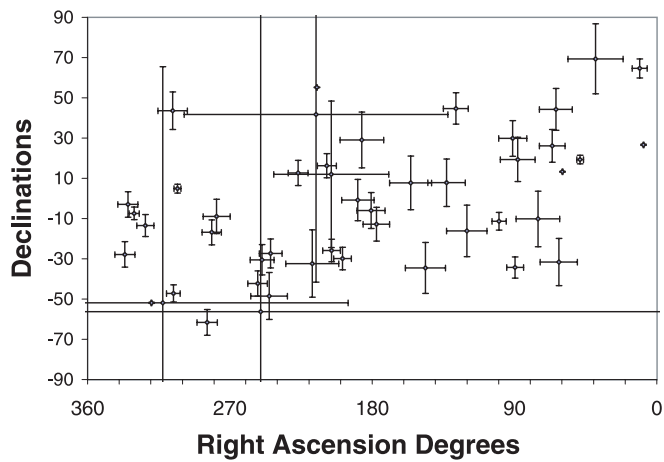

Fig. 7.—Terrestrial gamma flash in celestial coordinates with error bars. Some error bars are obviously too large and unrealistic for unknown reasons.

TGF sources in the Galactic plane center at maximal EGRET fluence; their correlations with important known TeV sources (Table 2) are displayed. Let us remark that the last discovered TeV source, 1ES 1426+428, associated with a BL Lac object at redshift $z = 0.129$ also correlates with the event in the TGF BATSE trigger 2955, making the TGF astrophysical nature more plausible than any random terrestrial lightening origin (Fig. 9). One should foresee that UPTAUs and HORTAUs must be correlated to geological sites of higher terrestrial densities (rock over sea), higher terrestrial crust elevation (mountain chains with less atmosphere opacity), as well as a higher terrestrial magnetic field. This correlation has been found (discussed in D. Fargion 2002, in preparation). Additional remarkable correlations occur with AGASA UHECR (Hayashida et al. 1999) inhomogeneities at the EeV energy band, as shown in Figure 10, as well as with most COMPTEL gamma sources toward $l = 18°$ in the Galactic plane. Some important locations of known Galactic and extragalactic sources (such as nearby quasi-stellar radio sources 3C 273 and 3C 279) are listed in Table 2 and are displayed in Figure 11 over the EeV AGASA map. Very recent (and the rarest) UHECR AGASA triplet clustering, near or above GZK energies and pointing toward BL Lac 1ES 0806+524, surprisingly finds a corresponding TGF event within its error box: BATSE trig-ger 2444. Also, two (of the four) additional UHECR AGASA doublets (2EG J0432+2910 and TEX 1428+370) are well correlated to TGF events (BATSE triggers 5317 and 2955). The present TGF–$\tau$ air shower identification could not be produced by UHE $\bar{\nu}_e$ charged current at $E_{\bar{\nu}_e} = M_W^2/2m_e = 6.3 \times 10^{15}$ eV; therefore, it stands for the UHE $\nu_\tau$-$\bar{\nu}_\tau$ presence. Consequently, it confirms $\nu_\mu \leftrightarrow \nu_\tau$ flavor mixing from far PSR or AGN sources toward Earth. The TGF–$\tau$ air shower connection may soon be verified and reinforced (or partially mystified) by the BATSE publishing of 28 missing TGF data (as well as future *GLAST* data): we foresee that BATSE TGFs are hiding additional directional imprints of UHE $\nu_\tau$ sources (maybe the missing Mrk 421 and Mrk 501 extragalactic sources).

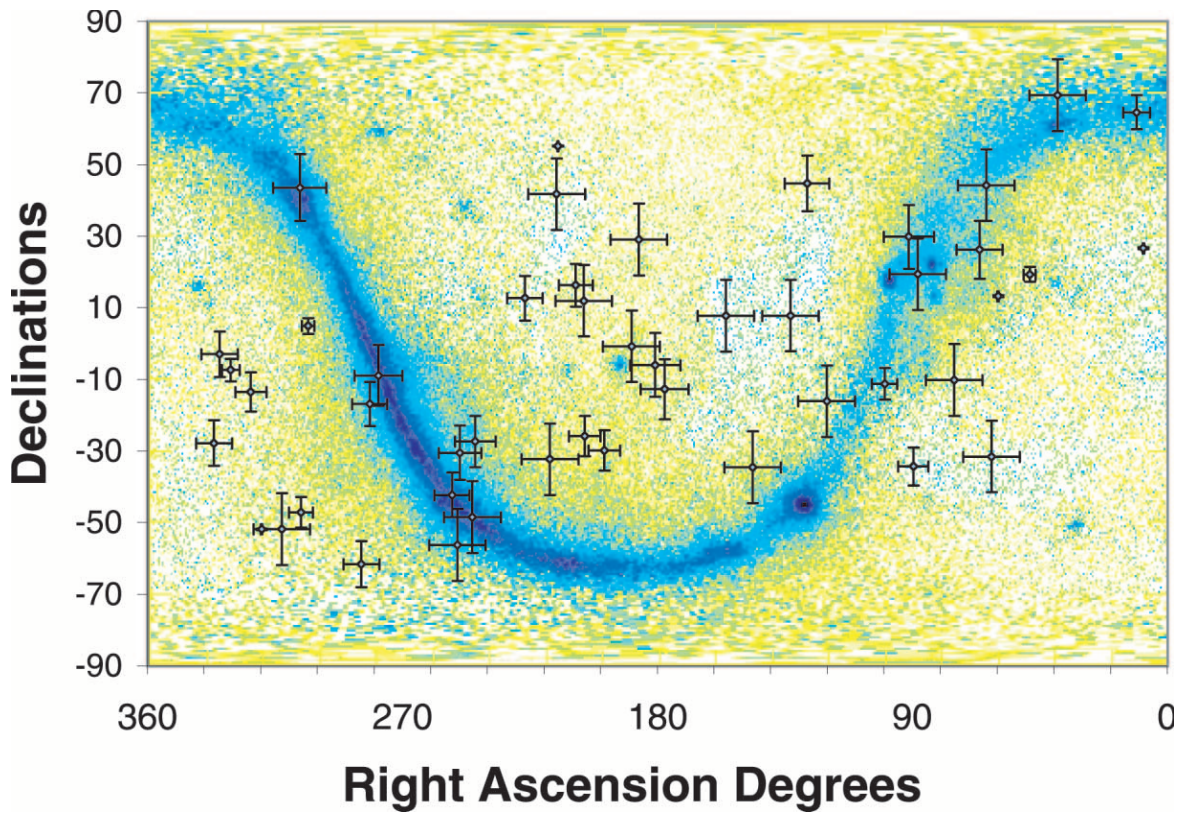

Fig. 8.—Terrestrial gamma flash after few error bar calibrations below maximal values, in celestial coordinates over the EGRET GeV diffused Galactic background.

## 4. CONCLUSIONS

In this article the UHE $\nu_\tau$ and $\tau$ propagation and the UHE $\tau$ air shower role have been analyzed. It has been shown that UHE horizontal and upward $\tau$ air showers within $E_\tau \gtrsim 3$ PeV may be observable and very probably have been already observed in upward BATSE TGF events. The distribution of seven (or eight) TGF events within a sample of 47 events at $\pm 3°$ along the Galactic plane may occur (Fig. 12) by chance (Poisson distribution) one time in 100 (or over 500), and it strongly favors the TGF–$\tau$ air shower connection. Similar conclusions arise from larger angle TGF data clustering from the Galactic plane ($\pm 10\%$; Fig. 13).

Moreover, the coincidence in direction (Fig. 14) and time (Fig. 15; repeaters) structure of some TGF events (shown by corresponding group labels in Table 3) makes their characteristic correspondence to active source very probable (blazing Galactic and extragalactic sources).[4] The observation of known sources toward the anti-Galactic center (Crab, PKS 0528, Geminga), toward the Galactic center (PSR 1706−44, PKS 1622−297, $\rho$ Oph, Sco X-1), and in the Galactic plane

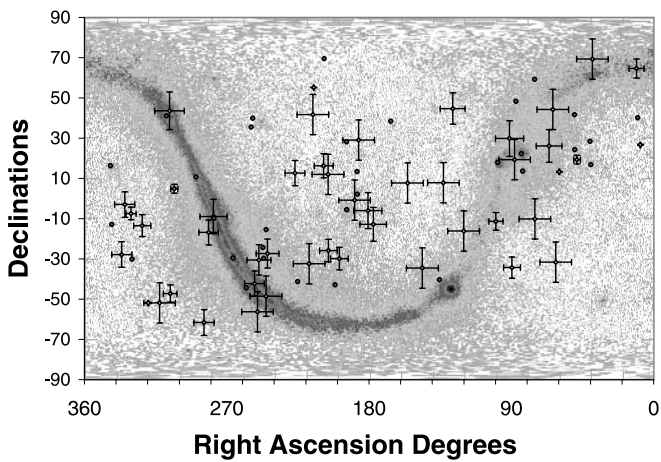

Fig. 9.—Terrestrial gamma flash data and relevant well-known TeV, X-ray, and gamma Galactic and extragalactic sources in red dots as in Table 2. [*See the electronic edition of the Journal for a color version of this figure.*]

---

[4] A natural question may arise: could the repeated TGF be a brief sequence of lightening? The answer (for all published BATSE data) is definitively no. Because the *CGRO* flies at the equator at 8 km s$^{-1}$, within a few hundreths of a second, it will be far from its original geophysical region.

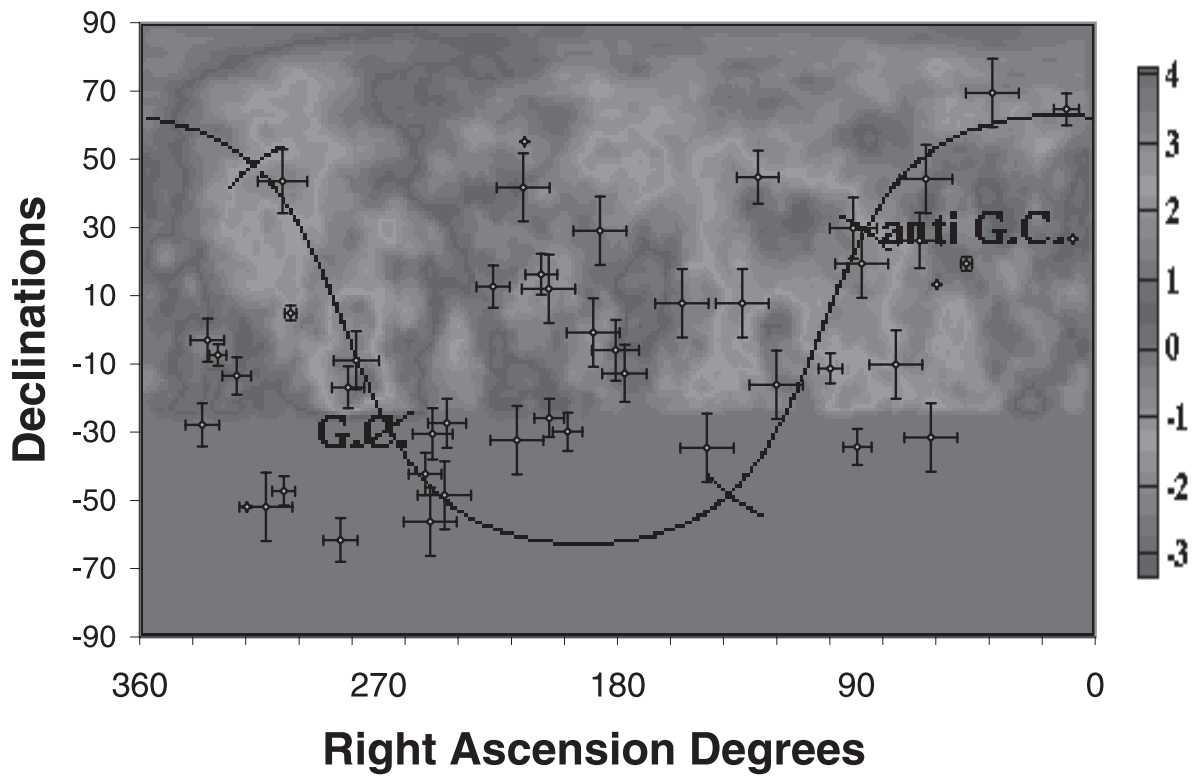

Fig. 10.—Terrestrial gamma flash in celestial coordinates over UHECR diffused data from AGASA cosmic rays at EeV energies. The clustering toward M17 and the Cygnus X-3 source should be noted. [*See the electronic edition of the Journal for a color version of this figure.*]

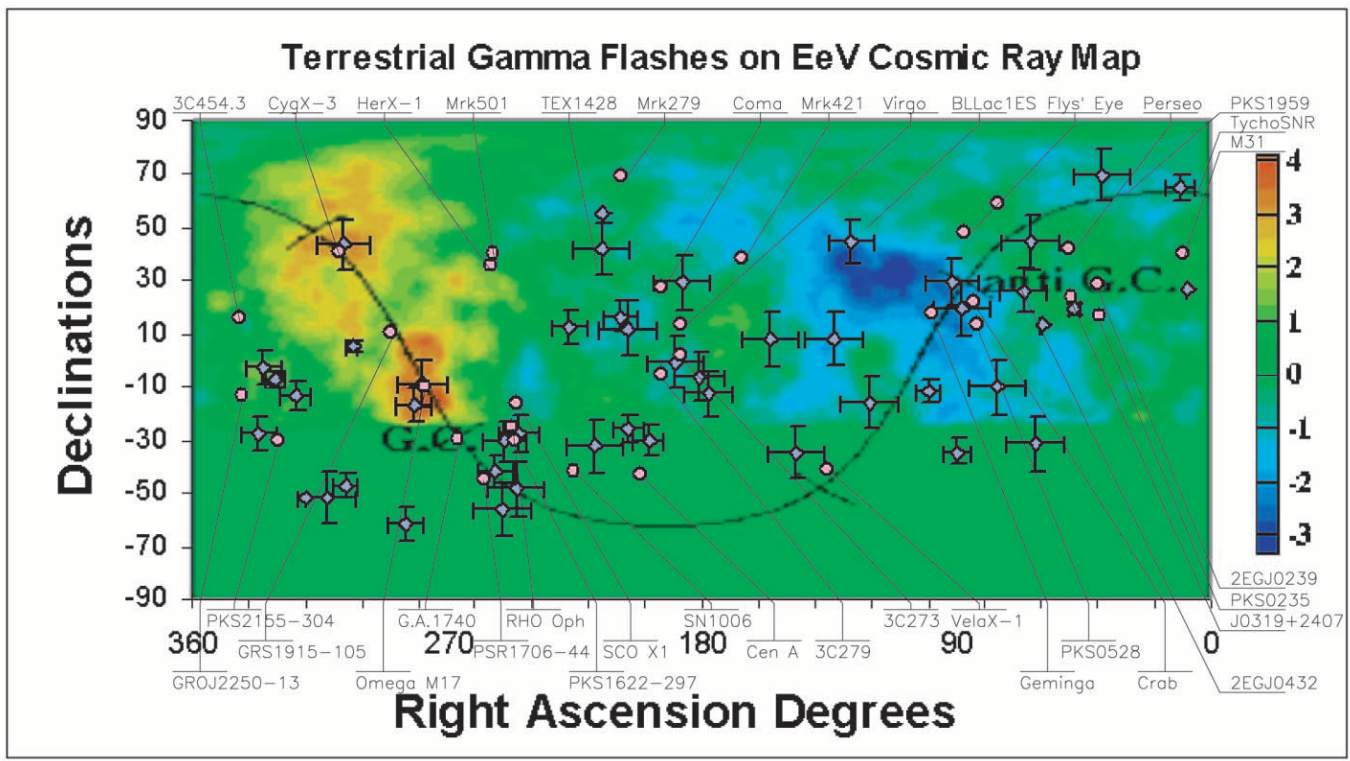

Fig. 11.—Terrestrial gamma flash in celestial coordinates over UHECR diffused data by AGASA cosmic rays at EeV energies with relevant known TeV, X-ray, and gamma Galactic and extragalactic sources in red dots as in Table 2.

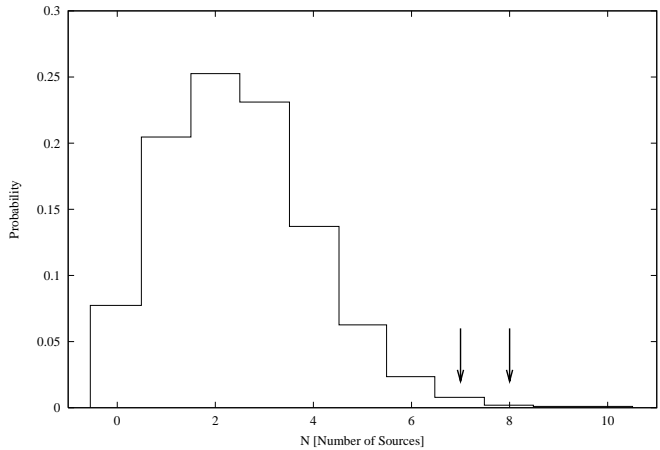

Fig. 12.—Poisson probability distribution (and simulation) to find by chance $N = 7\text{–}8$ TGF events within $\pm 3°$ of the Galactic plane.

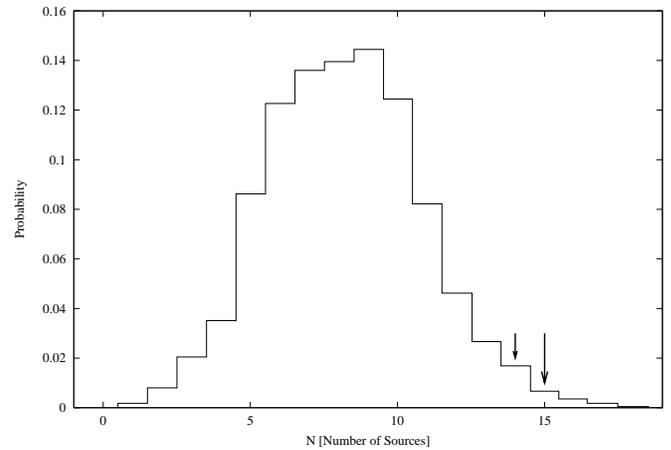

Fig. 13.—Poisson probability distribution (and simulation) to find by chance $N = 14\text{–}15$ TGF events within $\pm 10°$ of the Galactic plane.

(Cyg X-3, Cyg A, Tycho SNR, Omega M17) implies a minimal mixing distance and a consequently lower boundary distance ($\lesssim 4$ Kpc) in the oscillation $\nu_\mu \leftrightarrow \nu_\tau$ distance and a corresponding first upper bound to their square mass mixing: $\Delta m_{\nu_\mu \nu_\tau} \gtrsim 10^{-8}$ eV$^2$. Therefore, the need of UHE $\nu_\tau$'s calls for flavor mixing and $\nu_\tau$ mass, as Super-Kamiokande atmospheric neutrinos imply. A first rough consequence of the few observed events is the existence of a photopion opacity at both Galactic and extragalactic regions at a level comparable to (or below by an order of magnitude) model predictions (Stecker et al. 1991), at a rate of (one a month at PeV energy) $\sim (0.2\text{–}5) \times 10^{-13}$ events cm$^{-2}$ sr$^{-1}$. Because of the compelling role UHE $\nu_\tau$-$N$ interactions play in TGF events, we are also testing the same existence of the $\nu_\tau$ neutrino. We do not believe that nature would be so perverse as to mimic the inevitable signals by UPTAUs and HORTAUs into uncorrelated TGF events by just a finetuned chance. Therefore, an additional word, found within the 28 missing TGF data of the BATSE experiment, may be decisive. The UHE $\nu$ interaction at PeV–EeV energies corresponds to an invariant center of mass energy comparable to or above future LHC accelerators, and it offers an additional astrophysical laboratory for high-energy physics. The upward $\tau$ air shower, while opening a new UHE direction in neutrino astronomy, offers a deeper (UHE $\nu$'s are weekly interacting) view of the most violent cosmic accelerators.

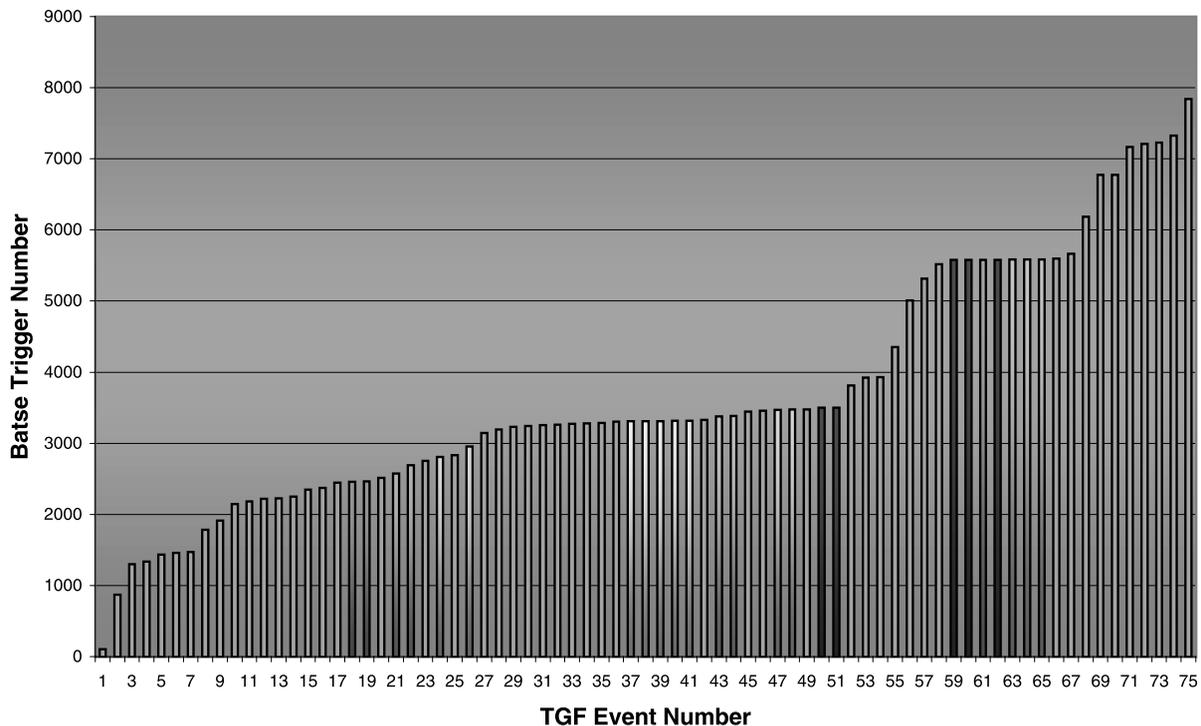

Fig. 14.—BATSE GRB trigger data from 1991 to 1999 as a function of the terrestrial gamma flash sequence. The different trigger setups—hard for channels 3+4—are the root of the two evident plateau growths in TGF rate; the group TGF events associate common arrival directions—Galactic center, anti-Galactic center, . . .—also related in time clustering as in Table 3. [*See the electronic edition of the Journal for a color version of this figure.*]

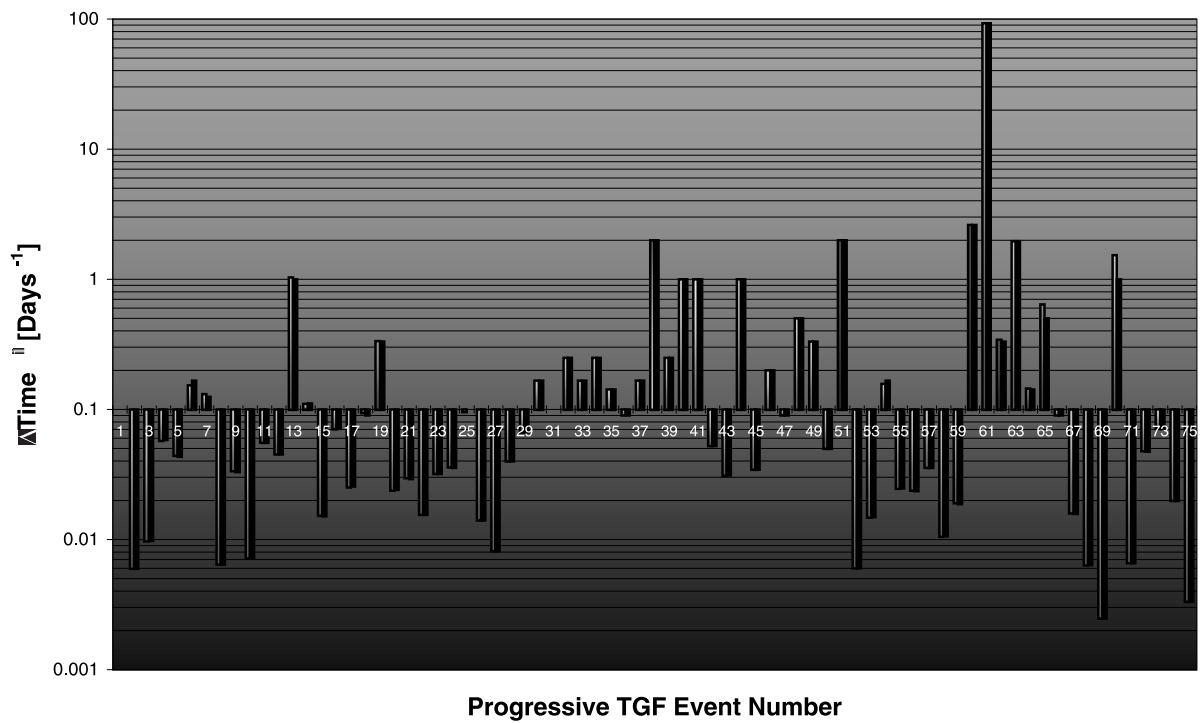

Fig. 15.—BATSE TGF rate from 1991 to 1999. The diagram shows the inverse of the time lag between to consecutive TGFs. The group TGF events associate common arrival directions (Galactic center, anti-Galactic center...) as in Table 3; the inverse time lapse (inverse time interval between two sequential TGF events) marks the probable repeater nature of a few TGF signals from the same arrival direction. [*See the electronic edition of the Journal for a color version of this figure.*]

This work originated at the Technion Institute (Fargion 1997).[5] The author thanks A. Dar and M. Bilenky for encouragement, F. Cesi for important suggestion on statistics, Barbara Mele and Tommasina Coviello for useful comments, suggestions, and support, and Pier Giorgio De Sanctis Lucentini for relevant graphical and computer tests and elaboration. The author is also grateful to Marco Grossi, Andrea Aiello, and Roberto Conversano. This work is based on the work of Bruno Rossi and Marcello Conversi, among others, concerning the discovery of cosmic rays and their opening of the lepton and quark physics; this paper is devoted to the memory of Bruno Pontecorvo, father of modern neutrino physics and astrophysics.

---

[5] After this article was submitted in 1997 and 2000, a list of articles considering the UHE $\tau$ role have appeared. Some of them have been included in the text; other important ones are Pasquali & Reno (1999), Athar, Jezabek, & Yasuda (2000a), Athar, Parente, & Zas (2000b), Feng et al. (2001), who referred to HORTAUs with the descriptive term "Earth-skimming UHE neutrinos," Bertou et al. (2002), and Cline & Stecker (2000).

TABLE 3
BATSE Terrestrial Gamma Bursts from 1991 to 1999

| Number | Group | Publication | Trigger Number | Date | Time (s) | R.A. (deg) | Decl. (deg) | $\Delta\theta$ (deg) | $\Delta\theta$ Geo (deg) | Triggers 3 + 4 |
|---|---|---|---|---|---|---|---|---|---|---|
| 1............ | | | 106 | 1991 Apr 22 | 2531.137856 | 99.74 | −11.31 | 4.42 | 18.298 | |
| 2............ | | | 868 | 1991 Oct 5 | ... | ... | ... | ... | ... | |
| 3............ | | | 1300 | 1992 Jan 15 | 47202.69786 | 217.85 | −32.34 | 16.71 | 50.007 | |
| 4............ | [1] | Not | 1334 | 1992 Feb 1 | 72419.98862 | ... | ... | ... | ... | |
| 5............ | [1] | Not | 1433 | 1992 Feb 24 | 36547.26786 | ... | ... | ... | ... | |
| 6............ | [1] | Not | 1457 | 1992 Mar 1 | 81250.81786 | ... | ... | ... | ... | |
| 7............ | | | 1470 | 1992 Mar 9 | 47072.13786 | 330.568 | −7.418 | 3.146 | 27.24 | |
| 8............ | | | 1787 | 1992 Aug 10 | 61515.16 | 305.76 | −47.18 | 4.26 | 20.965 | |
| 9............ | | | 1915 | 1992 Sep 9 | 28074.52 | 89.59 | −34.33 | 5.28 | 39.621 | |
| 10.......... | | | 2144 | 1992 Jan 24 | 54533.6 | 205.66 | −25.83 | 5.59 | 47.205 | |
| 11.......... | | | 2185 | 1992 Feb 11 | 53095.77 | 8.4 | 26.64 | 0.23 | 67.271 | |
| 12.......... | | | 2221 | 1993 Feb 5 | 55291.04 | 10.84 | 64.61 | 4.73 | 131.747 | |
| 13.......... | | | 2223 | 1993 Feb 6 | 52583.13 | 319.652 | −51.879 | 0.24893 | 51.343 | |
| 14.......... | | | 2248 | 1993 Feb 15 | 60330.14 | 59.71 | 13.27 | 0.25 | 151.511 | |
| 15.......... | | | 2348 | 1993 May 20 | 7337.69 | 281.52 | −16.86 | 6.15 | 59.37 | |
| 16.......... | | | 2370 | 1993 Jun 3 | 14440.54 | 252.45 | −42.27 | 6.21 | 24.21 | |
| 17.......... | | | 2444 | 1993 Jul 12 | 50022.45 | 127.12 | 44.73 | 7.83 | 35.895 | |

| Number | Group | Publication | Trigger Number | Date | Time (s) | R.A. (deg) | Decl. (deg) | Δθ (deg) | Δθ Geo (deg) | Triggers 3 + 4 |
|---|---|---|---|---|---|---|---|---|---|---|
| 18............ | [2] | | 2457 | 1993 Jul 23 | 18386.93 | 312.52 | −51.84 | 117.33 | 23.727 | |
| 19............ | [2] | | 2465 | 1993 Jul 26 | 16888.24 | 284.41 | −61.57 | 6.4 | 38.168 | |
| 20............ | | | 2516 | 1993 Sep 5 | 79940.98 | 244.25 | −27.38 | 7.19 | 32.963 | |
| 21............ | [3] | | 2573 | 1993 Oct 9 | 38648.63 | 205.96 | 11.98 | 36.38 | 35.732 | |
| 22............ | [3] | | 2692 | 1993 Dec 12 | 48679.66527 | 208.722 | 16.279 | 5.97512 | 17.985 | |
| 23............ | | | 2754 | 1994 Jan 12 | 49046.2 | 180.72 | −5.97 | 8.9 | 55.155 | |
| 24............ | [4] | | 2808 | 1994 Feb 9 | 22876.86 | 215.02 | 55.15 | 0.14 | 44.431 | |
| 25............ | | | 2835 | 1994 Feb 19 | 58464.88 | 323.46 | −13.5 | 5.49 | 35.827 | |
| 26............ | [4] | | 2955 | 1994 May 1 | 35887.92108 | 215.49 | 41.74 | 83.37 | 46.59 | |
| 27............ | [1] | Not | 3148 | 1994 Aug 31 | ... | ... | ... | ... | ... | |
| 28............ | [1] | Not | 3192 | 1994 Sep 25 | ... | ... | ... | ... | ... | On |
| 29............ | [1] | Not | 3233 | 1994 Oct 10 | ... | ... | ... | ... | ... | On |
| 30............ | [1] | Not | 3244 | 1994 Oct 16 | ... | ... | ... | ... | ... | On |
| 31............ | [1] | Not | 3258 | 1994 Oct 26 | ... | ... | ... | ... | ... | On |
| 32............ | [1] | Not | 3264 | 1994 Oct 30 | ... | ... | ... | ... | ... | On |
| 33............ | [1] | Not | 3274 | 1994 Nov 5 | ... | ... | ... | ... | ... | On |
| 34............ | [1] | Not | 3277 | 1994 Nov 9 | ... | ... | ... | ... | ... | On |
| 35............ | [1] | Not | 3285 | 1994 Nov 16 | ... | ... | ... | ... | ... | On |
| 36............ | [1] | Not | 3302 | 1994 Nov 27 | ... | ... | ... | ... | ... | On |
| 37............ | [5] | Not | 3309 | 1994 Dec 3 | ... | ... | ... | ... | ... | On |
| 38............ | [5] | Not | 3310 | 1994 Dec 3 | ... | ... | ... | ... | ... | On |
| 39............ | [6] | Not | 3313 | 1994 Dec 7 | ... | ... | ... | ... | ... | On |
| 40............ | [6] | Not | 3314 | 1994 Dec 8 | ... | ... | ... | ... | ... | On |
| 41............ | [6] | Not | 3315 | 1994 Dec 9 | ... | ... | ... | ... | ... | On |
| 42............ | | Not | 3331 | 1994 Dec 28 | ... | ... | ... | ... | ... | On |
| 43............ | [7] | | 3377 | 1995 Jan 29 | 23619.46508 | 132.94 | 7.82 | 11.79 | 33.696 | On |
| 44............ | [7] | Not | 3382 | 1995 Jan 30 | ... | ... | ... | ... | ... | On |
| 45............ | | Not | 3446 | 1995 Feb 28 | ... | ... | ... | ... | ... | On |
| 46............ | | Not | 3457 | 1995 Mar 5 | ... | ... | ... | ... | ... | On |
| 47............ | [8] | Not | 3470 | 1995 Mar 16 | ... | ... | ... | ... | ... | On |
| 48............ | [8] | Not | 3474 | 1995 Mar 18 | ... | ... | ... | ... | ... | On |
| 49............ | | Not | 3478 | 1995 Mar 21 | ... | ... | ... | ... | ... | On |
| 50............ | [9] | Not | 3500 | 1995 Apr 10 | ... | ... | ... | ... | ... | On |
| 51............ | [9] | Not | 3501 | 1995 Apr 10 | ... | ... | ... | ... | ... | On |
| 52............ | | | 3813 | 1995 Sep 22 | 4147.663844 | 177.363 | −12.77 | 8.38576 | 51.854 | |
| 53............ | | | 3925 | 1995 Nov 28 | 25539.1452 | 306.176 | 43.554 | 9.31987 | 52.484 | |
| 54............ | | | 3931 | 1995 Dec 4 | 55267.08114 | 189.102 | −0.774 | 10.23038 | 135.4 | |
| 55............ | | | 4355 | 1996 Jan 13 | 81867.29972 | 186.553 | 29.061 | 13.82847 | | |
| 56............ | | | 5006 | 1996 Feb 24 | 67333.92155 | 278.398 | −8.922 | 8.50541 | 21.476 | |
| 57............ | | | 5317 | 1996 Mar 23 | 55341.34553 | 66.203 | 26.187 | 8.12474 | 21.036 | |
| 58............ | | | 5520 | 1996 Jun 25 | 85244.19308 | 61.958 | −31.575 | 11.75859 | 46.283 | On |
| 59............ | [10] | | 5577 | 1996 Aug 17 | 13701.40916 | 88 | 19.354 | 11.00976 | 32.485 | On |
| 60............ | [10] | | 5578 | 1996 Aug 17 | 46631.65578 | 75.155 | −10.188 | 13.80963 | 17.464 | On |
| 61............ | [11] | | 5579 | 1996 Aug 17 | 47563.04128 | 155.715 | 7.75 | 13.28786 | 37.48 | On |
| 62............ | [10] | | 5582 | 1996 Aug 20 | 39893.02514 | 91.151 | 29.822 | 8.91121 | 29.862 | On |
| 63............ | [12] | | 5583 | 1996 Aug 20 | 83982.24111 | 250.51 | −56.233 | 469.78418 | 116.46 | On |
| 64............ | [12] | | 5587 | 1996 Aug 27 | 74000.92911 | 245.215 | −48.473 | 11.66136 | 54.871 | On |
| 65............ | [12] | | 5588 | 1996 Aug 29 | 35537.63997 | 249.631 | −30.47 | 7.50174 | 29.902 | On |
| 66............ | | | 5598 | 1996 Sep 9 | 42071.07306 | 336.451 | −27.808 | 6.35217 | 43.103 | |
| 67............ | | | 5665 | 1996 Nov 11 | 6460.705047 | 38.713 | 69.385 | 17.44419 | 73.637 | |
| 68............ | | | 6185 | 1997 Apr 16 | 71107.68123 | 226.787 | 12.669 | 6.2585 | 16.835 | |
| 69............ | | | 6773 | 1998 May 22 | 76751.80148 | 63.817 | 44.243 | 10.37695 | 61.128 | |
| 70............ | | | 6777 | 1998 May 23 | 46630.20105 | 334.484 | −2.996 | 6.34613 | 31.81 | |
| 71............ | | | 7168 | 1998 Oct 21 | 57752.06039 | 146.259 | −34.542 | 12.68023 | 51.353 | |
| 72............ | | | 7208 | 1998 Nov 11 | 44176.31316 | 198.714 | −29.859 | 5.58169 | 27.523 | |
| 73............ | | | 7229 | 1998 Nov 25 | 44884.85698 | 303.202 | 4.903 | 2.16981 | 54.592 | |
| 74............ | | | 7325 | 1999 Jan 14 | 53731.02702 | 120.195 | −16.118 | 12.77933 | 13.653 | |
| 75............ | | | 7844 | 1999 Nov 8 | 17993.61694 | 48.525 | 19.338 | 2.09525 | 23.785 | |

NOTE.—All BATSE terrestrial gamma burst data from 1991 to 1999. Sources with common "Group" designations have a possible common source origin. The group TGF events associate common arrival directions (such as the Galactic center, anti-Galactic Center) associated also in time clustering; the date, time, celestial coordinates, error bars, and TGF-Earth center angle are listed below; hard trigger setup trigger periods (channels 3 + 4) have a (triggers 3 + 4) side label. They mark a higher rate TGF activity visibly correlated with two different plateaus in Fig. 14 corresponding to higher TGF acceptance. The last three TGF events during 2000, triggers 8006, 8083, and 8108, confirm but did not change the general result.

## APPENDIX A

### $\tau$ AIR SHOWER TIMING

Upward and downward air showers *are not symmetric* events at all because of the different atmosphere densities at sea level and high altitude. Indeed, at sea-level $\tau$ air showers hold for just a microsecond, but at high levels $\tau$ decays produce millisecond showers.

The arrival time of gamma air showers (bremsstrahlung photons) is ruled by the last atmosphere distance at which the gamma emission has originated (while being nearly unabsorbed). The mean energy deposition profile in the air shower is given by a common gamma distribution:

$$\frac{dE}{dt} = \frac{E_0 b (bt)^{a-1} e^{-bt}}{\Gamma(a)} , \qquad (A1)$$

where the adimensional shower depth distance $t = x/X_0$ and the adimensional energy $y = E/E_c$ are well-known variables. The characteristic critical energy $E_c$ (Rossi 1964) is, in air, around 100 MeV. The air shower maxima occurs at an adimensional depth $t_{max} = a - 1/b$, while the characteristic shower distance $X_s \equiv X_0/b$, with $b \simeq 0.5$, is $X_s \simeq 2X_0$ [note that $t_{max} = (a-1)/b \simeq (\ln y + \frac{1}{2})$ defines $a$ for photon-induced cascades]. Naturally, the radiation length $X_0$ is the same for upward and downward air showers. However, the corresponding length distances are very different because altitudes (sea level and high altitude) have extremely different densities. The air density decreases with respect to the sea-level height with altitude $z$ as $\rho = \rho_0 e^{-(z/h_0)}$, with $h_0 \simeq 8.55$ km. If one considers the sea-level case $X_0 = 36.6$ g cm$^{-2}$ and that the radiation distance is $X_0 = 304.2$ m, the shower length is $X_s \simeq 608.4$ m, and the corresponding shower scale time is (as it is well known and observed in common downward air showers) $t_s = X_s/c \simeq 2$ $\mu$s; this result is obvious. If now we consider upward $\tau$ air showers arising on the top atmosphere altitude, then the same $X_0 = 36.6$ g cm$^{-2}$ corresponds, in a more diluted upward atmosphere, to distances $X_0' \simeq 22$ km and in a first approximation to a shower length $X_s' \simeq 44$ km, leading to $t_s' \simeq 75 t_s \sim 0.15$ ms. Additional time dilutions must be considered for the arrival nadir angle $\theta$, $t_s' \sim 0.2/\cos\theta$ ms, and geomagnetic Larmor precession of relativistic electrons. This delay is maximal for horizontal $\tau$ air showers. More precisely, the $\tau$ air shower timing is related to the total distance from the earliest atmosphere's last scattering ($X_0 \sim 36.6$ g cm$^{-2}$ and $h \simeq 22$ km) up to the BATSE satellite height ($\sim$500 km); therefore, the maximal upward $\tau$ air shower extends up to $t_s' \sim 500/c \cos\theta$ km $> 2$ ms. The exponential density decay in the upper atmosphere makes most of the bremsstrahlung radiation generated at the lowest altitudes (altitudes of tens of kilometers), implying a fast rise of the gamma flash within a few tens of milliseconds (such as the observed TGF ones) even if the gamma signal must also extend up to a few milliseconds, as is indeed observed in TGFs. Different $\tau$ air shower channels (Table 1) and their consequent bifurcation may lead to rapid TGF millisecond timing modulations, as in the observed ones.

## APPENDIX B

### TEN QUESTIONS AND ANSWERS ON THE ULTRAHIGH $\nu$ AND $\tau$ ROLE

1. Why does the time structure of terrestrial gamma flashes (due to UHE $\nu$ and UHE $\tau$ upward air showers) show a millisecond duration, while the observed downward air showers are only microseconds long?

Because upward and downward air showers (at sea level and high altitudes, respectively) are not symmetric processes because of the extreme difference (nearly 2 orders of magnitude) in the corresponding air density environment. The different shower scale times are derived from these differences (see Appendix A).

2. Why is sprite lightening not a good explanation of the TGF events?

Because their characteristic observed duration holds between 10 and 1000 ms (Wright 2000), while the TGF rising time structure is as short as 0.1 ms (2–4 orders of magnitude below).

Also, the most probable sprite event rate (one event every 30 s on Earth; Wright 2000) occurring on the whole Earth (area $\sim 5 \times 10^8$ km$^2$) may hit the observation area of the *CGRO* satellite ($\lesssim 3 \times 10^6$ km$^2$), leading to an expected TGF rate $\dot{N}_{TGF} \simeq \dot{N}_{sprites}(A_{GRO}/A_{Earth}) \simeq 2 \times 10^{-4}$ s; on the contrary, the observed TGF event rate (10 yr, 78 events) on average is $3 \times 10^{-7}$ s, leading to a discrepancy of nearly 3 orders of magnitude.

3. Why might the low energetic PeV $\tau$ ($10^{15}$–$10^{17}$ eV) air showers explain the observed TGF gamma fluence ($\Phi_\gamma \sim 10^{-2}$ cm$^{-2}$; $F_\nu \sim 10^{-9}$ ergs cm$^{-2}$ $\sim 10^3$ eV cm$^{-2}$)?

Because the UHE $\tau$ air showers ($E_\tau \gtrsim 10^{15}$–$10^{17}$ eV) are beamed jets spread in a narrow area of $10^{10}$–$10^{12}$ cm$^2$, leading to the above observed gamma fluxes at a rate for the main predicted UHE $\nu$ spectra model, comparable to the observed one.

4. Why do the upward $\tau$ air shower spectra agree with the hard TGF spectra?

Because the electromagnetic cascade by $\tau$ air showers ends with hard bremsstrahlung photons.

5. Why is the idea to look for UHE $\nu$ and UHE $\tau$ showering from mountains or large nadir angles from Earth so recent?

Because the key penetrating role of UHE $\tau$ in matter has been noted only recently (Fargion 1997) along with the more penetrating role of UHE $\nu_\tau$ (Halzen 1998; Gandhi et al. 1998) and the last strong evidence of a $\nu_\mu$-$\nu_\tau$ flavor mixing derived by Super-Kamiokande. All of these arguments linked with the unstable behavior of $\tau$ and its decay in air were leading us to the present $\tau$ air showers as a powerful tool for UHE $\nu$ discover.

6. Why do the $\tau$ air shower detectors have a sensitivity to UHE $\nu$ exceeding the cubic kilometer ice or water detector under construction?

Because other detectors trace mainly muons, seeking their long track through their Cerenkov radiation photons. However, the photon transparency lengths in water or ice ($L \lesssim 20$ m) imply a very frequent array of elements (in principle, $50^3 \simeq 10^5$, but at least 5000 phototubes are necessary). Moreover, the total radiation losses in the Cerenkov photons in cubic kilometers are only a tiny fraction [$\sim 10^{-8}(E_\mu/\mathrm{PeV})$] of the primary UHE $\nu$ energy. Therefore, the effective detection volume for each optical tube is small ($V_{\mathrm{eff}} \sim 2.5 \times 10^{-6}$ km$^{-3}$), and the total number of released photons in the cubic kilometer track is limited [$N_{\mathrm{op}} \sim 2 \times 10^7 (E_\mu/\mathrm{PeV})$]. Nuclear and/or electromagnetic showering in ice or water are useless because their huge diffused isotropic signals are severely bounded by Landau & Pomeranchuck (1953)—Migdal (1956) effect at a few meters, and they have no directionality and energy calibration (indeed, these photon flashes saturate the phototubes).

On the contrary, $\tau$ air showers produced at a distance of a few kilometers from a mountain chain or upward to mountains, balloons, or satellites release huge amounts (nearly all) of the primary $\tau$ (and primary UHE $\nu$) energy: $N_{\mathrm{opt}} \simeq 10^{12}(E_\tau/\mathrm{PeV})$; $N_\gamma(\langle E_\gamma \rangle \sim 10$ MeV$) \simeq 10^8(E_\tau/\mathrm{PeV})$; $N_{e^-e^+} \simeq 2 \times 10^7(E_\tau/\mathrm{PeV})$; $N_\mu \simeq 3 \times 10^5(E_\tau/\mathrm{PeV})^{0.85}$. Therefore, the ratio of the two used signal fluxes in cubic kilometers and in $\tau$ air showers is at least 5 orders of magnitude in favor of $\tau$ showering.

Moreover, in the wide observable area ($\gtrsim \pi R_M^2 \gtrsim 3 \times 10^4$ m$^2$), $\tau$ air shower spread implies a smaller number of needed detectors. For a nominal mountain valley of 10 km in length and 1 km in height, 200 detector elements at a reciprocal distance of a 2–4 Moliere radii may be enough. For upward $\tau$ air showers toward a 1 km mountain, ten (or a few tens) detectors may be needed. For UPTAUs toward a satellite, one unique detector of a few square meters, such as BATSE or *GLAST*, is just enough.

7. Why do $\tau$ showers in water or ice lose their primordial directionality information, while $\tau$ air showers keep memory of it?

Because the Cerenkov showering in water/ice occurs within a wide angle, $\theta_w = \{2[1 - (1/n\beta)]\}^{1/2} \simeq 42°$, while the corresponding relativistic showering in air occurs within a much narrower Cerenkov cone, $\theta_a = [2(n - 1)]^{1/2} \simeq 1°33$, comparable with the same air shower jet cone ($\sim 1°$).

8. Why are even the widest Auger detectors seeking horizontal air showers not competitive with $\tau$ air showers (at a comparable area)?

Because the same huge air target volume above the Auger detector ($\sim 50$ km$^3$) at horizontal arrival directions plays a disruptive role (in reducing the UHE $\nu$ showers effects) through its air opacity (Cillis & Sciutto 2001). Indeed, at low zenith vertical arrival angles, Auger detectors are ruled by common hadronic cosmic-ray shower noise; at larger zenith angles, $90° > \theta > 70°$, the severe air opacity to gamma shower signals (a slant depth $\gtrsim 2000$ g cm$^{-2}$) leads to a flat suppressed muonic shower component nearly 3 orders of magnitude smaller than the main peak electromagnetic shower. Moreover, Auger detectors record events at the highest ($E_\nu \gtrsim 10^{18}$–$10^{19}$ eV) energies at which the expected UHE $\nu$ flux is suppressed (with respect to PeV $\nu$), by at least 3 orders of magnitude. A possible improvement is the idea to move and locate the Auger detector much nearer to the Andes Mountain chain to observe their lateral $\tau$ showering from the mountain chain (Fargion et al. 1999a).

9. Why concentrate on $\tau$ air showers mainly at $10^{15}$–$10^{17}$ eV?

Because the combined interaction probability growth with energy, the Earth opacity, the UHE $\tau$ length, the confined atmosphere height of a few kilometers, and the same atmospheric UHE $\nu$ noise lead to such a narrow "opportunity" energy window: $5 \times 10^{17}$ eV $\gtrsim E_\tau \gtrsim 10^{15}$ eV.

10. What experimental configurations for $\tau$ air showers are necessary?

Because of the complete analogy between $\tau$ hadronic and nucleon or nuclear showers, the needed detector configurations are just comparable to known air shower detectors of cosmic rays at the knee (Fowler et al. 2001).

CASA-BLANCA, DICE, CASA-MIA, Akeno, and Tibet are among the known ones. The expected $\tau$ air shower flux will be (for comparable primary $\nu$ and cosmic-ray spectra at PeV) nearly 5 orders of magnitude smaller than common hadronic ones, but their presence behind a mountain a few degrees below the edge will be rare (a couple of days for a PeV $\nu$ flux comparable to the cosmic-ray flux) but much above the noise threshold and the angular uncertainty ($\pm 0°5$) reconstructed zenith angle (Ave et al. 2000). For a UHE $\nu$ flux $\sim E^{-2}$ much below the cosmic-ray flux (at the ratio $\eta = \Phi_\nu/\Phi_{\mathrm{CR}}$, with $\eta \simeq 10^{-2}$ to $10^{-3}$; Berezinsky & Ginzburg 1990; Stecker et al. 1991), the expected event rate for a CASA-BLANCA squared kilometer detector is $150\eta(E/\mathrm{PeV})^{0.363}$ events yr$^{-1}$. A wider detection area, like 1% of Auger (50 km$^2$), would observe tens of events at $10^{16}$ eV yr$^{-1}$ even if $\eta \sim 10^{-3}$. For a more detailed event rate derivation, see § 2. A wide detector area in an array located on a high mountain inclined down toward Earth at large zenith angle (or nadir angle) below the horizon would inspect larger areas and detect even more events at PeV energy windows up to EeV energies at the far horizon. Note that because energetic EeV $\tau$ lengths exceed 50 km before decay, the horizontal air shower could originate deep in the horizontal long atmosphere depth, contrary to comparable EeV hadronic showers severely suppressed by air opacity.

## APPENDIX C

### $\tau$ INTERACTION LENGTHS

Let us briefly describe the estimate of $\tau$ versus $\mu$ interaction lengths. First, we remind the reader of the simplest $\tau$ range by boosted decay flight $R_{\tau_0}$:

$$R_{\tau_0} = c\tau_\tau \gamma_\tau = 4.902 \text{ km}\left(\frac{E_\tau}{10^8 \text{ GeV}}\right) . \tag{C1}$$

Let us now also consider the length due only to $\tau$ bremsstrahlung radiation losses $R_{R_\tau}$. The radiation length $b_\tau^{-1}$ will play a role in defining the $\tau$ range by the general energy loss equation:

$$-\frac{dE_\tau}{dx} = a(E_\tau) + b_\tau(E_\tau)E_\tau \,, \tag{C2}$$

where $a$ and $b$ are slow energy variable functions for ionization and radiation losses, respectively. For UHE $\tau$'s and muons, the ionization losses may be neglected. The asymptotic radiation length $b_\tau^{-1}$ at high energies $E_\tau \gg 10^{15}$ eV is related to the corresponding muon length by this approximated relation derived by a classical bremsstrahlung formula and is scaled for the two different lepton masses leading, after integration (Fargion 1997), to an extreme $\tau$ $R_{R_\tau}$ range:

$$R_{R_\tau} = 1033 \text{ km} \frac{5}{\rho_r}\left\{1 + \frac{\ln\left[(E_\tau/10^8 \text{ GeV})(E_\tau^{\min}/10^4 \text{ GeV})^{-1}\right]}{\ln 10^4}\right\}. \tag{C3}$$

However, these large distances are neglecting additional updated terms due to pair production and photonuclear losses (Becattini & Bottai 2001; Dutta et al. 2001) and electro-weak deep inelastic scattering (Fargion 1997). In order to find an analytical solution to the most general $\tau$ and updated energy losses, we define here the final energy fraction $x_i \equiv E_{\tau f}/E_{\tau i}$. Let us then also estimate the $\tau$ length boosted by its relativistic Lorentz factor as a function of the same final energy fraction $x_i$: $R_{\tau\text{Nucl}_\rho} = x_i c(E_{\tau i}/m_\tau)\tau_\tau$. Solving the radiation energy loss equation to find the $\tau$ length and equating to its boosted length, we found for each initial $E_{\tau i}$ the corresponding energy fraction $x_i$ and the consequent $\tau$ range described in Figure 1; the transcendental equations to be solved for each $x_i$ corresponding to initial energy $E_{\tau i}$ are

$$\frac{\ln(1/x_i)}{\rho_r[b_{0\tau} + b_{1\tau}\ln(E_{\tau i}/E_{0\tau})]} = 492 x_i \frac{E_{\tau i}}{E_{0\tau}} \,, \tag{C4}$$

where the minimal energy constant $E_{0\tau}$, above which our phenomenological law may be applied, is $E_{0\tau} \equiv 10^{14}$ eV; the factor 492 is the corresponding $\tau$ decay length at that energy expressed in centimeters. The adimensional relative density factor $\rho_r$ is unity for water. Finally, $b_{0\tau}$ contains comparably constant terms from pair production and photonuclear interaction, while $b_{1\tau}$ is due only to photonuclear effects: $b_{0\tau} = 2.5 \times 10^{-7}$ cm$^{-2}$ g$^{-1}$ and $b_{1\tau} = 6.5 \times 10^{-8}$ cm$^{-2}$ g$^{-1}$. The fractional energies values $x_i$ are estimated and drawn in the figure for all $\tau$ ranges in water and rock. For comparison, a simplest (overestimated) muon range is also shown (Fargion 1997; Lipari & Stanev 1991). It is evident that above few EeV energies $\tau$ ranges overcome muons, making such UHE $\tau$ neutrinos the easiest neutrinos to observe. The ratio among the $\tau$ and muon range is comparable. The above expression underestimated the $\tau$ lengths, and a very similar solution (which is at most a few percent above the previous lengths) may be derived solving the following comparable transcendental equations:

$$\frac{\ln(1/x_i)}{\rho_r[b_{0\tau} + b_{1\tau}\ln(E_{\tau i}x_i/E_{0\tau})]} = 492 x_i \frac{E_{\tau i}}{E_{0\tau}} \,. \tag{C5}$$

Let us remind you of the energy losses due to electro-weak interaction ranges $R_{\text{Weak}_\rho}$, which become dominant at extreme (GZK) energies and suppress both $\tau$'s and muons as well as neutrino propagation in matter (see Fig. 1):

$$\frac{1}{\sigma N_A \rho_r} \simeq \frac{2.6 \times 10^3 \text{ km}}{\rho_r}\left(\frac{E_\tau}{10^8 \text{ GeV}}\right)^{-0.363}. \tag{C6}$$

Finally, let us notice the possible severe opacity for both $\tau$'s and neutrinos because of new physics (extradimension gravity) interactions at TeV. The corresponding $\tau$ range is $R_{\text{New}} = (\sigma_{\text{New}} N_A \rho_r)^{-1}$, which equals

$$16.6\left(\frac{E_\tau}{10^8 \text{ GeV}}\right)^{-1}\left(\frac{E_{\text{New}}^{\text{TeV}}}{10^3 \text{ GeV}}\right)^{-4}\frac{\text{km}}{\rho_r}. \tag{C7}$$

This opacity will mask UPTAUs, but it will amplify by 2 orders of magnitude the HORTAUs behind a mountain chains. In Figures 1 and 3 we summarized the corresponding curve for the $\tau$ and UHE neutrino ranges in different density matters.